\documentclass[12pt,preprint]{aastex}

\begin{document}

\title{A Synthetic Spectrum and Light Curve Analysis of the Cataclysmic Variable IX Velorum\footnotemark[1]}
\footnotetext[1]
{Based on observations made with the NASA/ESA Hubble Space Telescope, obtained at the
Space Telescope Science Institute, which is operated by the Association of Universities 
for Research in Astronomy, Inc. under NASA contract NAS5-26555, and the NASA-CNES-CSA
{\it Far Ultraviolet Explorer}, which is operated for NASA by the Johns Hopkins University
under NASA contract NAS5-32985} 


\author{Albert P. Linnell$^2$, Patrick Godon$^3$, Ivan Hubeny$^4$, Edward M. Sion$^5$,
and Paula Szkody$^6$}

\affil{$^2$Department of Astronomy, University of Washington, Box 351580, Seattle,
WA 98195-1580\\
$^3$Department of Astronomy and Astrophysics, Villanova University,
Villanova, PA 19085\\
visiting at the Space Telescope Institute, Baltimore, MD.\\
$^4$Steward Observatory and Department of Astronomy,
University of Arizona, Tucson, AZ 85721\\
$^5$Department of Astronomy and Astrophysics, Villanova University,
Villanova, PA 19085\\
$^6$Department of Astronomy, University of Washington, Box 351580, Seattle,
WA 98195-1580\\
}

\email{$^2$linnell@astro.washington.edu\\
$^3$godon@stsci.edu\\
$^4$hubeny@as.arizona.edu\\
$^5$edward.sion@villanova.edu\\
$^6$szkody@astro.washington.edu\\
}

\begin{abstract}

Spectrum synthesis analysis of $FUSE$ and $STIS$ spectra for 
the cataclysmic variable IX Vel shows that it is possible to achieve a close synthetic spectrum fit
with a mass transfer rate of ${\dot M}=5{\times}10^{-9}M_{\odot}/{\rm yr}^{-1}$ and a modified
standard model temperature profile.	The innermost four annuli of the accretion disk model, 
extending to $r/r_{\rm wd}{\approx}4$, are
isothermal; beyond that limit the temperatures follow the standard model.

A light synthesis fit to K band photometry requires shallow eclipses of the accretion disk rim and 
secondary star limb. The geometry constrains
the orbital inclination to $i=57{\pm}2{\arcdeg}$.
The synthetic light curve 
requires a vertically extended accretion disk rim, beyond that 
predicted by gravitational
equilibrium, to shadow the secondary star. The enhanced vertical extension is consistent with recent 
MHD predictions for CV accretion disks.

Matching differing observed heights of alternate K band light curve maxima requires a warm rim region downstream of
the intersection point of the mass transfer stream with the accretion disk rim. The temperature
of the warm region is inconsistent with expectation for a bright spot associated with a rim shock.

\end{abstract}




\section{Introduction}

Cataclysmic variables (CVs) are semi-detached binary stars in which a late main sequence star 
loses mass onto a white dwarf (WD) by Roche lobe overflow \citep{w1995}. 
In non-magnetic systems the mass
transfer stream produces an accretion disk with mass transport inward and angular momentum 
transport outward driven by viscous processes. The accretion disk may extend inward to the WD; the
outer boundary extends to a tidal cutoff limit imposed by the secondary star in the steady state case.
If the mass transfer rate is below a certain limit, the accretion disk is unstable
and undergoes brightness cycles (outbursts), and if above the limit, the 
accretion disk is stable against outbursts;
the latter case objects are called nova-like (NL) systems.

NL systems are of special interest because they are expected to have an accretion disk radial temperature
profile	given by an analytic expression \citep[eq.5.41]{fr92}(hereafterFKR) thereby defining
the so-called standard model. IX Vel is one such NL system.

IX Vel was discovered as a CV by \citet{gar1982}
in connection with their MK-UBV survey of southern OB stars \citep{gar1977}.
They reported colors similar to UX UMa, light flickering of order $0.1^{\rm m}$ on
a timescale of minutes, with larger variations on the order of years, and extremely broad, shallow H
absorption lines with emission cores. The light variations in NL systems are thought to
arise from variations in the mass transfer rate \citep[][\S4.3.1]{w1995}.
\citet{war1983} presented additional spectrophotometry, with a radial velocity curve
based on mean values of the first three Balmer lines, together with a preliminary
estimate of the orbital period, $0.187^{\rm d}$. 
\citet{war1984} presented photometric observations showing that
the photometric magnitude varied erratically between $m_{pg}=9^{\rm m}.1$ and $10^{\rm m}.0$
between Dec. 1963 and Jan. 1973. Their IR photometry showed no obvious eclipse effects,
setting a limit to the inclination of roughly $65-70\arcdeg$. \citet{gar1984} provided 
further optical spectral
data, confirmed that IX Vel is subject to flickering on a time scale of minutes, and derived
an estimated mass transfer rate of $7{\times}10^{-10}M_{\odot}{\rm yr}^{-1}$.
\citet{eggen1984} confirmed the photometric variation and suspected doubling in 
the H${\alpha}$ and H${\beta}$ emission lines.
\citet{war1985} used high speed photometry to show that IX Vel is variable at the $0.001^{\rm m}$
level with a non-constant flickering period ${\sim}25^{\rm s}$.

\citet{sion1985} was first to model the system; he used $IUE$ spectra and
called attention to the presence of a wind from the system. Sion showed that the energy
distribution is consistent with a mass transfer rate between $10^{-8}$ and 
$10^{-9}M_{\odot}{\rm yr}^{-1}$; his analysis showed that IX Vel is a
NL system.

\citet{haug1988} obtained JHK photometry of the system and showed that there is
a low-amplitude IR light curve with an amplitude of a few times 0.01, with a
deeper minimum at phase 0.0 and slightly unequal successive light maxima.

\citet{beu1990} (hereafter BT) detected narrow optical H and He emission lines 
from the secondary star
in IX Vel. The narrow lines were in antiphase to somewhat broader emission lines
that varied in phase with the WD and accretion disk, and in turn were superimposed
on broad absorption troughs. Radial velocities for the emission components, and a
more accurate orbital period ($0.1979292^{\rm d}$), determined
component masses for the system ($M_1=0.80M_{\odot}$, $M_2=0.52M_{\odot}$) and a 
value for the orbital inclination ($i=60{\arcdeg}$). The model also determined a
mass transfer rate of $7.9{\times}10^{-9}M_{\odot}/{\rm yr}^{-1}$ and a distance
of 95 pc. by Bailey's method (see BT for details).
BT attributed the emission line excitation to irradiation by the primary or the
accretion disk, supporting a suggestion by \citet{hess1989}. 
The BT model has been the accepted model to the present time. 

\citet{long1994}, hereafter L94, used HUT to obtain a spectrum of IX Vel in the wavelength range
830-1860\AA. They modeled the continuum with a sum of stellar synthetic spectra to
represent the individual annuli, and assumed that the resulting disk spectrum was
the only contributor to the system synthetic spectrum. L94 obtained the important 
results	that: (1) a standard model accretion disk (FKR) does not provide an
accurate fit to the HUT spectrum, and (2) the quality of fit is sensitive to the
adopted interstellar reddening of the observed spectrum; they state that the WD
in the system likely has a $T_{\rm eff}$ in the range 50,000-75,000K.
By contrast, \citet{sw1991} had found a good standard model fit to spectral energy distribution (SED)
data listed by \citet{h1987}.

\citet{mauche1991} used high resolution $IUE$ spectra to study the wind associated
with IX Vel. Further studies on this topic have been by \citet{prinja1995} and
\citet{hartley2002}.

\citet{kubiak1999} presented phase-resolved spectroscopy which provides evidence for
an accretion disk rim bright spot contribution to the system spectrum. 
\citet{polidan1990} detected no extreme-UV flux from IX Vel with {\it{Voyager}}, but
\citet{teeseling1995} detected X-rays from IX Vel with ROSAT. The latter authors 
studied ROSAT and EUVE observations of IX Vel and showed that there is an orbital
modulation of X-ray flux which probably excludes a boundary layer origin but might
arise from a rim bright spot.
They also found that the
x-ray flux is more than a thousand times smaller than the far UV flux, and that the x-ray
flux can be modeled with a single optically thin spectrum.

A motivation for this study is that	the
$FUSE$ spectra embrace the peak of the IX Vel SED near 1050\AA, providing an improved constraint
on a synthetic spectrum fit.
In addition, the Hipparcos parallax \citep{perry1997} places IX Vel
within the Sun's immediate neighborhood and permits an absolute calibration of the mass transfer 
rate from the secondary star.

\section{Interstellar reddening}

The L94 result emphasizing the importance of interstellar reddening leads us to consider
this topic with care.

\citet{gar1984} propose E(B-V)=0.03  for IX Vel and a distance of approximately
100 pc, while
\citet{Bohn1982} list E(B-V)=0.02 based on ESO $IUE$ spectra; by contrast 
\citet{ladous1991} lists E(B-V)=0.0, based on $IUE$ spectra SWP18578 and
LWR14655, while 
\citet{bruch1994} list E(B-V)=0.01. 
\citet{mauche1988} studied narrow interstellar absorption lines on $IUE$ spectra
to determine H I column densities near the Sun, determined a value of
$20{\times}10^{19}{\rm cm}^{-2}$ in the direction of IX Vel, and 
note that \citet{ver1987}, 
determined a
rough upper limit of E(B-V) of 0.04 (and a preferred value of E(B-V)=0.0) based on a study 
of the 2200\AA~``bump"
in $IUE$ spectra. 
L94 note that, if the gas to dust ratio along the line of
sight to IX Vel is typical, the reddening corresponding to the column density
of H I inferred by \citet{mauche1988} implies E(B-V)$\approx$0.004 mag.
L94 found that the best-fitting model spectrum depends on 
both the adopted E(B-V) and the
adopted mass transfer rate from the secondary star. With the distance to IX Vel
set at 95 pc., from BT, the best-fitting model spectrum corresponded to E(B-V)=0.07
and a mass transfer rate of $1.3{\times}10^{-8}M_{\odot}/{\rm yr}^{-1}$. This
mass transfer rate was more than 10 times higher than for the best-fitting model
spectrum if E(B-V)=0.0. L94 found that, on adopting the mass transfer rate derived by BT, the
best-fitting model spectrum corresponded to ${\rm E(B-V)}=0.03$. 
Based on the available data
we believe the most likely E(B-V)=0.01, which we adopt and confirm {\it \'{a} postiori}.

\section{The $FUSE$ and $STIS$ spectra}

The observations log is presented in Table~1 . 

\subsection{The FUSE Spectrum}  

The spectrum of IX Vel was obtained 
through the 30"x30" LWRS Large Square Aperture, in HISTOGRAM mode. 
The data were processed with CalFUSE version 3.0.7.
The main change from previous versions of CalFUSE is that now the data
are maintained as a photon list (the intermediate data file - IDF) throughout
the pipeline. Bad photons are flagged but not discarded, so the user can examine
the filter, and combine data without re-running the pipeline. 
The spectrum actually consists of 12 exposures of about 600 sec each, and  
these exposures, totaling 7336 sec of good exposure time, were combined (using the IDF file).  
Consequently,  
the FUSE spectrum has a duration of about 0.44 of the period of IX Vel.
The spectral regions covered by the FUSE 
spectral channels overlap, and these overlap regions are then used to
renormalize the spectra in the SiC1, LiF2, and SiC2 channels to the flux in
the LiF1 channel. We combined the individual channels to create a
time-averaged spectrum with a linear $0.1$ \AA\, dispersion, weighting
the flux in each output datum by the exposure time and sensitivity of the
input exposure and channel of origin. 
In this manner we produced a final spectrum that covers the
full {\it{FUSE}} wavelength range $905-1187$ \AA . 

\subsection{The HST/STIS spectrum}  

The STIS spectrum was processed with CALSTIS version 2.22. 
The HST/STIS spectrum of IX Vel was obtained in TIME TAG 
operation mode using the FUV MAMA detector.  
The observations were made through the 0.1x0.03 aperture 
with the E140M optical element. The STIS spectrum has a 
duration of 1750 sec or about 0.1 the period of IX Vel.
The STIS spectrum, centered on the wavelength 1425\AA\ , 
consists of 42 echelle spectra that we extracted and combined
to form a single spectrum from about 1160 \AA~to 1710 \AA.  
At very short wavelengths the STIS spectrum is very noisy and we
disregard the region $\lambda < 1160$ \AA;   
towards longer wavelengths the echelle 
spectra do not overlap and five gaps are apparent
around $\lambda \approx $1634\AA , 1653\AA , 1672\AA , 1691\AA , and
1710\AA . 

Figure~1 shows the FUSE and the STIS
spectra of IX Vel.

\subsection{The Lines} 

The FUV spectra of accreting WDs in CVs exhibit absorption 
and emission lines from the WD, the accretion disk, the  
hot component (possibly the inner disk or boundary/spread layer)
as well as sharp absorption lines from circumbinary
material and/or the ISM. Additionally,
terrestrial contamination is also usually observed, especially in
FUSE spectra. 

The main characteristic of the {\it{FUSE}} spectra of CVs 
is the broad Ly$\beta$ absorption feature due to the large 
gravity (log $g\approx 8$) and temperature of the WD, 
while the Ly$\alpha$ absorption feature is observed in the STIS spectral
range. In the disk these features are usually broader and smoothed out unless 
the inclination of the system is very low and the disk is almost face-on.
At higher temperatures, as the continuum rises in the shorter wavelengths,
the higher orders of the Lyman series also become visible in the
FUSE range. 

Examination of the {\it{FUSE}} spectrum reveals
that the continuum is very slightly affected by deep and sharp 
ISM H\,{\sc i} absorption towards shorter wavelengths, 
extending to the hydrogen cut-off around 915\AA\ . We annotated 
the position of these hydrogen lines under the first panel of Figure~1. 
Other lines which are usually attributed to the ISM and/or to air
contamination are the O\,{\sc i} lines and the low order Nitrogen 
lines N\,{\sc i} \& N\,{\sc ii}. All these lines are observed in
the FUSE spectrum of IX Vel. Since these lines are sharp and are observed
at their theoretical wavelengths they are not associated with the
source, though they could be from circumbinary material ejected
from the system. Of particular interest are the Si\,{\sc ii}, 
S\,{\sc ii}, and C\,{\sc ii} lines observed in the STIS spectral range.
These lines are usually observed as broad absorption lines originating
from the WD atmosphere. Here, these lines are sharp, and, like the 
H\,{\sc i}, O\,{\sc i}, N\,{\sc i}, and N\,{\sc ii} lines, they are
located at their theoretical wavelengths. This seems to indicate that
the circumbinary material in the line of sight was ejected from the
WD/accretion disk. The sharp Si\,{\sc iii} 1206.5\AA~line is at its 
theoretical wavelength and appears to be circumbinary.
Not as obvious, but also apparently coming from
the circumstellar gas, are the sharp lines of 
Si\,{\sc iv}, and C\,{\sc iv} superposed on the broad
absorption lines.  

All the broad absorption lines in the FUSE spectrum
(N\,{\sc iv}, S\,{\sc vi}, C\,{\sc iii}, N\,{\sc iii}, O\,{\sc vi},   
H\,{\sc i}, S\,{\sc iv},...)  and in the STIS spectrum 
(C\,{\sc iii}, Si\,  N\,{\sc v}, Si\,{\sc iv}, and  
C\,{\sc iv}) have the same shape 
with a sharp steep red wing and a much broader blue wing.
The H\,{\sc I}, C\,{\sc iv}, and He\,{\sc ii}
lines do have a pronounced P Cygni profile while all the remaining
broad absorption lines present only a hint of possible 
red shifted emission. This is an indication that the FUSE and STIS
spectra were obtained at about the same orbital phase (see e.g.
Hartley et al. 2002).  

In Table~2 and Table~3 we identify all the absorption lines of the FUSE
and STIS spectra. The wavelength tabulations are the theoretical wavelengths.

\subsection{Combining the spectra} 

We consulted the {\it{AAVSO}} (American Association of Variable Stars 
Observers) light curve generator to check the magnitude  of IX Vel 
at the time of the observations. However, the AAVSO data did not provide
data on the specific days the observations were made but a few days 
before and later. The source appeared to vary erratically between about 
magnitude 9.3 and 9.9 ($9.6 \pm 0.3$) in April 2000 (when the FUSE 
spectra (\S3.1) were obtained (April~15)) and between about magnitude 9.5 and 10.1 
($9.8 \pm 0.3$) in July of the same year (when the 
STIS spectrum was obtained (July~19)). (\citet{hartley2002}, their Figure~1 and Table~2,
assign this observation to August~19 instead of July~19.)
To within the 0.3 magnitude uncertainty, 
the FUSE spectrum and the STIS spectrum were at about the same magnitude
($\approx$9.6 versus $\approx$9.8). 
In addition, from the flux level of the FUSE and STIS spectra it is clear
that the system was in the same state and with the same mass accretion rate. 
Finally, from a visual inspection of the absorption
lines it is clear that the spectra were obtained 
at about the same orbital phase. 

Before combining the FUSE and STIS spectra, we stress that 
(i) the system was in the same state at the two times of observation; 
(ii) the shapes of the spectra (continuum
and lines) are similar (where the spectra are not too noisy); 
(iii) the flux level in the FUSE and HST/STIS spectra do not differ 
more than about 5\% (the FUSE and HST/STIS are very different instruments and
we do not expect their flux level to match exactly); (iv) the 
spectra have been obtained at about the same binary
phase (this is important since IX Vel is a high inclination 
system for which the observing time is significantly shorter than
the orbital period).  
We have adopted a divisor of 1.03, applied to the $STIS$ spectrum,
in placing the two spectra on the same plot. We have also removed flux entries from both spectra
which were either 0.0 or negative. 

\section{The analysis program:BINSYN}

Our analysis uses the program suite BINSYN \citep{linnell1996}; a recent paper \citep{linnell2007}
describes its
application to a CV system in detail. Briefly, an initial calculation produces a set of annulus
models for a given WD mass, radius, and mass transfer rate. This calculation uses the program
TLUSTY \citep{hubeny1988,hl1995}. 
The program considers the radial and vertical structure of the disk independently.
The radial structure is based on the standard model (FKR) and so follows the prescribed relation
between local $T_{\rm eff}$ and the annulus radius. The vertical structure is solved,
self-consistently, as described by \citet{hubeny1990} and \citet{hh1998}.
The set of annulus models covers the accretion disk from its innermost (WD) radius to 
just short of its outermost 
(tidal cutoff) radius. Our final model, discussed below, adopts an outer accretion disk $T_{\rm eff}$
which the final and penultimate annulus models bracket.

Table~4 lists properties of annuli calculated with TLUSTY(v.202)
for a mass transfer rate of $5.0{\times}10^{-9}M_{\odot}/{\rm yr}^{-1}$. 
The control file to calculate a given annulus requires a radius of the WD in units of $R_{\odot}$.
All of the annuli used the radius of a zero temperature WD for a homogeneous 
carbon Hamada-Salpeter
$0.80M_{\odot}$ model from \citet{panei2000}.
All of the annuli are H-He
models, and the models through $r/r_{\rm wd,0}=20.00$ are converged LTE models. The remaining annuli are
LTE-grey models (see the TLUSTY Users Guide for an explanation).
Each line in the table 
represents a separate annulus. The column headed by $m_0$ is the column mass, in ${\rm gm}~{\rm cm}^{-2}$, 
above the central plane.
The column headed by $T_0$ is the upper boundary temperature in degrees Kelvin, with the upper boundary 
specified as that 
level with a
column mass of $1.0{\times}10^{-5}{\rm gm}~{\rm cm^{-2}}$ lying between it and the next inward tabular level. 
The entire annulus has a total of 70 levels.
Compare with the annulus
$T_{\rm eff}$ value, which references the local temperature at a Rosseland optical depth of ${\approx}0.7$. 
The columns headed by $z_0$ and $N_e$ are respectively the upper boundary
height (cm) above the central plane and the electron density (${\rm cm}^{-3}$) at the upper boundary. 
The $z_0$ value in the final row of Table~4 determines the rim semi-height.
The column headed
by ${\rm log}~g$ is the log gravity in the $z$ direction at a Rosseland optical depth of ${\approx}0.7$. 
The column headed by ${\tau}_{\rm Ross}$
is the Rosseland optical depth at the central plane. We call attention to the fact that the annuli are optically
thick to the outer radius of the accretion disk.

Following calculation of annulus models for assigned $M_{\rm wd}$ and $\dot{M}$,
program SYNSPEC(v.48)~\footnote{http://nova.astro.umd.edu} \citep{hlj1994} is used to produce a 
synthetic spectrum for each annulus.
We adopted solar composition for all synthetic spectra. The synthetic spectra include contributions
from the first 30 atomic species, assumed to be in LTE.

SYNSPEC optionally produces either flux values or intensities at a specified number of zenith angles
for the individual annulus synthetic spectra. The spectra, produced in BINSYN by interpolation among the
SYNSPEC spectra, require intensity data for integration over the visible object. If the SYNSPEC output
is in flux units, BINSYN associates a limb darkening value to produce intensity values at required
directions. We tested both the intensity and flux SYNSPEC options for the simulations described below,
using a limb darkening value of 0.6 
in connection with the 
flux output;
the difference between the two types of system synthetic spectra was nearly undetectable.

BINSYN models the complete CV system, including the WD, secondary star, accretion disk face,
and accretion disk rim as separate entities. The model represents phase-dependent and 
inclination-dependent effects, including eclipse effects and irradiation effects, on all of the system objects. The
representation of the stars requires polar $T_{\rm eff}$ values and gravity-darkening exponents.
BINSYN represents the accretion disk by a specified
number of annuli, where that number typically is larger than the number of TLUSTY annulus
models. The accretion disk $T_{\rm eff}$ profile may follow the standard model, but, alternatively, the profile
may be specified by a separate input file. The contributions of the individual BINSYN annuli to the
BINSYN representation of the accretion disk face follows by interpolation within the set of
SYNSPEC spectra for the TLUSTY annuli. The $T_{\rm eff}$ of the accretion disk rim ties to the
outermost annulus $T_{\rm eff}$ via the theory by \citet{hub1991}. 

BINSYN has provision to represent a bright spot on the rim face, with specified local $T_{\rm eff}$,
angular extent, and $T_{\rm eff}$ taper, upstream and downstream, relative to the specified
bright spot location. The output synthetic spectra include the separate contributions of the two stars,
the accretion disk face, the accretion disk rim, and the composite system spectrum.

BINSYN also can calculate a synthetic light curve for specified wavelength (see \S6.2 and the following
sections).

\section{System parameters adopted initially}

Initially adopted parameters from BT (Table~1) included $M_{\rm wd}=0.80M_{\odot}$, 
$M_{\rm sec}=0.52M_{\odot}$, P$=0.1939292^{\rm d}$, and 
$i=60\arcdeg$. \citet{panei2000} lists a WD radius of $9.8459{\times}10^{-3}R_{\odot}$
for a $0.80M_{\odot}$ homogeneous zero temperature Hamada-Salpeter carbon model WD.
In our subsequent study of the observed spectra we correct the radius for the adopted WD $T_{\rm eff}$;
the contribution of the WD to the system synthetic spectrum is appreciable. In the
immediately following study of the K band light curve the relatively insignificant and constant contribution
of the WD is absorbed in the normalization of system light to its maximum value. Consequently,
we use the accretion disk as calculated for	the zero temperature WD in the light curve study.
The secondary star mass and the results of \citet{kb2000} determine
a secondary star spectral type of M2. This spectral type and 
Table~15.7 of \citet{cox2000} determines a polar $T_{\rm eff}$ 
of 3500K for the secondary star. This is the same temperature adopted by BT (see our Table~5).
Note that a main sequence class V star of the same mass
would have a spectral type slightly earlier than M0 (Table~15.8 of \citet{cox2000}) and
a polar $T_{\rm eff}$ of 3900K.
In calculating irradiation of the secondary star we have
adopted a bolometric albedo of $A_{\rm s}=0.6$ \citep{claret2001}; $A_{\rm wd}=1.0$.
We have used standard values for gravity darkening exponents of ${\beta}_{\rm wd}=0.25$
and ${\beta}_{\rm s}=0.08$.

A number of studies have considered the tidal cutoff boundary, $r_d$, of accretion disks
(\citet{pac1977}; \citet{pp1977}; \citet{wh1988}; \citet{scr1988}; \citet{wk1991}; \citet{g1993}).
These authors agree on $r_d{\approx}0.33D$, where $D$ is the separation of the stellar components.
We adopt this expression for the tidal cutoff radius of the accretion disk.

The Hipparcos parallax \citep{perry1997} is $10.38\pm0.98~{\rm mas}$, 
corresponding to a distance
of 96.3 pc which we adopt for the system distance.

\section{An IX Velorum system model}

\subsection{An initial model}

Starting with the parameters of the previous section, we 
added an explicit model of
the accretion disk. 
Using TLUSTY annulus models for $M_{\rm wd}=0.80M_{\odot}$ (still zero temperature)
and a mass transfer rate of $8.0{\times}10^{-9}M_{\odot}/{\rm yr}^{-1}$ (BT), we calculated
a standard model system synthetic spectrum. The rim semi-height, $0.085R_{\odot}$, followed from the 
TLUSTY model for
the outer annulus, and the radius of the accretion disk was $0.56R_{\odot}$, the tidal
cutoff radius. We divided the synthetic spectrum by $8.8299{\times}10^{40}$, the scaling
factor corresponding to the Hipparcos distance of 96.3 pc, for comparison with the unreddened
observed spectrum.
The synthetic spectrum flux was appreciably too high over the entire spectral range of the
$FUSE$ plus $STIS$ spectra
for our adopted reddening
of ${\rm E(B-V)}=0.01$.
A standard model synthetic spectrum with ${\dot M}=5.0{\times}10^{-9}M_{\odot}/{\rm yr}^{-1}$
had a much improved overall fit. In this model the rim semi-height is $0.080R_{\odot}$, equal to
the height of the outermost TLUSTY (equilibrium) model. 
Although the synthetic spectrum was slightly too blue, and the flux level too high, we
anticipated improving both the spectral gradient and the flux level by modifying the accretion disk 
temperature profile from the standard model.	

The inclination $i=60\arcdeg$ (BT) is a maximum value, for a test model, to avoid deep eclipses
which are not observed.	In fact, this model has grazing eclipses; we tested the effect on the synthetic	spectrum
of
reducing $i$ to $54\arcdeg$, which barely avoids eclipses for the current model. Because of the reduced inclination,
flux from the $54\arcdeg$ model was both too large and too blue. Reducing the
mass transfer rate has a very small effect on the continuum gradient, so we modified the
accretion disk temperature profile. The hot inner
annuli have the strongest effect on the spectral gradient. We reduced the temperatures of the
innermost four annuli to 30,000K and produced a fairly close fit to the observed spectral gradient and flux level, 
still at $i=54\arcdeg$.
The outer annuli make small contributions to the synthetic spectrum, so the synthetic spectrum is
relatively insensitive to changes in the temperature profile in the outer accretion disk. However,
the light curve results are sensitive to the temperature structure of the outer accretion disk.
Consequently, at this stage we calculated a K band (2.2$\mu$) synthetic light curve for comparison
with the observational data by \citet{haug1988}.

\subsection{Treatment of the disk rim and a K band synthetic light curve} 

We used the Dexter facility of the Astrophysics Data System (hereafter ADS) to
extract the observed points from Haug's Fig.7. Our program PGCAC, within the BINSYN suite,
evaluates irradiation of the secondary star due to the WD, the accretion disk faces, and
the accretion disk rim. The effects are calculated separately, permitting an evaluation
of the independent effects. The combined effects, with the adopted rim semi-height of $0.080R_{\odot}$,
produced an Algol type light curve, 
with a single light maximum per orbital cycle near phase 0.5, and in strong
conflict with the observations. This situation had been anticipated by BT, and they had
proposed an irradiation model in which the rim height is high enough to shield most
of the secondary from the WD. 

This new constraint requires replacing the annulus heights calculated by TLUSTY;
it also means that the previously calculated synthetic spectra
must be recalculated.
We set the semi-height of the rim to an  empirical (i.e., not theoretically-based) and
iteratively-determined height of
$0.20R_{\odot}$; this height barely shadows the poles of the secondary from the WD, but
unavoidably guarantees that shallow eclipses of the secondary star and 
the accretion disk take place, even for an inclination of $54\arcdeg$. The secondary star is partially
irradiated by the accretion disk face regions more remote than the WD.
We discuss the light curve analysis below, but with the system geometry now approximately
fixed we first focus on production of a suitable synthetic spectrum.
We found that very good fits to the $STIS + FUSE$ spectra are possible by adjusting the
$T_{\rm eff}$ values of the four innermost annuli, with $i$ between $54\arcdeg$ and 
$60\arcdeg$, the ${\dot M}=5.0{\times}10^{-9}M_{\odot}/{\rm yr}^{-1}$ temperature
profile for the remainder of the accretion disk, and reddening corresponding 
to $E(B-V)=0.01$.
Figure~2 portrays
the system projected on the plane of the sky at orbital phase 0.0, for an orbital inclination of
$57\arcdeg$. 

The following text describes a further adjustment of the accretion disk temperature profile, 
but it is first necessary
to return to the light curve analysis. \citet{haug1988} attempted a simulation of the light
curve based on the assumption that the secondary star and the WD both have $T_{\rm eff}=3550$K(!),
that irradiation of the secondary is by an accretion disk with $T_{\rm eff}{\leq}5100$K(!), and
that the bolometric albedo of the secondary is 1.0. There was no evaluation of shadowing effects
nor details concerning the irradiation calculation. 
Haug obtained a difference in depths of the minima by setting the secondary star bolometric albedo
to 1.0. Although the calculated light curve fitted the observations fairly well, the Haug
parameters are grossly unrealistic.	Haug was unable to simulate the difference in the alternate
light maxima.

Shadowing of the secondary by the increased rim height has an important effect on the temperature
profile of the secondary star. In the absence of any irradiation, the $T_{\rm eff,s}$ of the
secondary star photospheric 
point, essentially coincident with the L1 point (hereafter 'point'), is 618K
in our model (the calculated point gravity is small but not zero). Although the
point receives no irradiation
from the WD or faces of the accretion disk, there is irradiation by the accretion disk rim. 
This irradiation produces a point $T_{\rm eff}$ value of 3119K. Irradiation of the
secondary star by the rim affects the calculated light curve.

The light loss at the successive minima is a complex function of the temperature
distribution on the object undergoing eclipse and the eclipsed areas. Only a segment of the
accretion disk rim undergoes eclipse, and the rim temperature depends strongly on the system
model. In our model the top of the rim barely shadows the poles of the secondary star from the WD.
BINSYN does not
have the capability to simulate semi-transparent regions, but we assume the top of the rim
permits passage of enough photons to produce the observed secondary component emission lines,
originating near the secondary star poles.
We use the theory of \citet{hub1991} to relate the rim temperature to the temperature
of the outermost annulus, which the standard model assumes to be in Keplerian motion. 
Briefly, the outer region of the accretion disk is not expected to be precisely in Keplerian motion,
the outer boundary is expected to be approximately semi-toroidal
in cross-section and to have an effective temperature given by
\begin{equation}
T_{\rm eff,rim}/T_{\rm eff,Kep}=(8q/9)^{1/4},
\end{equation}
where
\begin{equation}
q=\frac{4}{\pi}\frac{1+(h/2)}{[1+(2/{\pi})h][1+2h]}, ~h=\frac{H}{R},
\end{equation}
where $H$ is the semi-height of the rim and $R$ is the radius of the rim upper edge, measured
in the central plane. 
The temperature of the outermost annulus depends on the mass transfer
rate; for the adopted ${\dot M}$, and the standard model, that temperature is 5659K, giving $T_{\rm eff,rim}=5052$K.
That temperature is below the $T_{\rm eff}$ of the outermost annulus in Table~4 because the rim radius is
slightly larger than the radius of the last calculated annulus. 
Note that the temperature in question
is below the temperature nominally adopted for transition to the hot state (${\approx}6300$K \citep{smak1982}). 
We defer any discussion of this 
point and proceed to
calculate a K band light curve for the current system parameters.

For light curve purposes, we represent the radiation from a given area segment on any of the system objects by a
black body at the local $T_{\rm eff}$. The light intensity at polar angle $\gamma$ relative to the local surface
normal and at wavelength $\lambda$ is
\begin{equation}
I_{\lambda}(\gamma)=B_{\lambda}[1-u_1(1-{\rm cos}({\gamma}))-u_2(1-{\rm cos}({\gamma}))^2],
\end{equation}
where $B_{\lambda}$ is the black body intensity. The quadratic limb darkening coefficients, $u_1$ and $u_2$, are
from \citet{wr1985}. Before presenting the light curve calculations it is necessary to consider the different
heights of the alternate light maxima.

\subsection{A bright spot}

\citet{kubiak1999} use their artificial spectrum-template generator \citep{poj1997} to produce phase-resolved 
artificial spectra for comparison with observed spectra. Their three-component model, with one component a
bright spot, is a close match to the summed observed spectrum. They derive an orbital phase of 0.75 for
the bright spot. With respect to the observed light curve,
we assume the difference in successive 
maximum light levels																	
arises from the interaction of the transfer stream and the accretion disk outer boundary region.
Figure~3 shows a plan view of the system. The mass transfer stream follows the theory of \citet{ls1975}.
The stream enters the rim at an angle
of $62\arcdeg$ relative to a tangent line at the impact point.
If the bright spot were centered at the impact point the higher light maximum would occur at orbital phase 
0.90, contrary to observation. 
A warm downstream rim region centered at orbital phase 0.75, as derived by \citet{kubiak1999},
produces a good fit to the observed light maxima. This location is close to the orbital phase 0.8 found in
other systems \citep{szkody1992,livio1993}.
We assume the location and $T_{\rm eff}$ of the warm rim region roughly represents
that dissipation signature. Figure~4 shows the system at orbital phase 0.75; the dark rim area indicates the
warm region. In this figure the secondary star is approaching the observer and the warm region is at its maximum
phasewise exposure.	The light curve is not strongly sensitive to the phasewise location of the spot; variation
of 0.05 produces little effect on the calculated light curve.
Our software cannot produce a bright spot more restricted in height on the rim than shown. The dark rim area,
Figure~4, does not represent the tapered temperature profile at the spot beginning and end. A spot
model with an oval appearance would produce a photometric signature differing negligibly from the model
presented here.

\subsection{Synthetic light curve fits and implications}

Figure~5 presents the synthetic K band light curve overplotted on the photometric observations. The observations 
(\citet{haug1988}) are
for 1984 January 16-23 (crosses) and 1987 January 12-16 (asterisks). The plotted points represent observations binned
in 0.1 phase intervals, with initial numerical values measured relative to the mean for the time interval in question.
The original Haug observations are in differential magnitudes. Haug estimates the individual observations have a 
0.03 mag.accuracy. The binned observations are more accurate than this.
The change in the light curve between the two dates is believed to be real.
We have applied a -0.025 magnitude correction to the
observations for optimum fit to the normalized theoretical light curve. Note that the ordinate is in flux units,
not magnitudes.

The depths of the light minima are functions of the orbital inclination; it is of interest to determine the sensitivity
of the minima to the inclination. Figure~6 shows the synthetic light curve for $i=54\arcdeg$, and Figure~7 is
the plot for $i=60\arcdeg$. All other parameters have remained fixed. Based on the three plots, we argue that the
best fit is for $i=57\arcdeg$ and that the estimated accuracy is $\pm 2\arcdeg$. The light
curve geometric requirement for a shallow eclipse of the accretion disk rim also confirms the tidal 
cutoff radius of the accretion disk used in this analysis.

The observations for the two epochs show differences that may reflect bright spot changes arising from changes in the mass
transfer rate. For this reason it is inadvisable to attempt an optimized light curve fit. 

We call attention to the
greater primary minimum phase width than presented by the Figure~5 model, for both sets of observations. 
The ``break" in the
synthetic light curve on both the ingress and egress phases coincides with the beginning and end of geometric
eclipse of the accretion disk rim. The observed light reduction prior to that geometric eclipse indicates that 
eclipse of a light source already is in progress.
It is tempting to associate the light reduction with eclipse of a light source associated with the
wind known to be present (and discussed below). Actual demonstration of this speculation as a physical possibility
would require extended analysis beyond the scope of this paper.
As can be seen from Figure~2, the limb of the secondary star would start to eclipse a wind region before beginning eclipse
of the accretion disk rim. Inclusion of an additional light source (similar to ``third light" in classical binary
systems)
would require renormalization of the system light unit and so would reduce the calculated depth of the phase 0.0
light minimum, improving the synthetic light curve fit to the observations.

\subsection{The rim $T_{\rm eff}$ and the final system model}

The $T_{\rm eff}$ values of the outer annuli in the current light curve model are below the transition value 
to the hot state, which
has a minimum $T_{\rm eff}$ of approximately 6300K \citep{smak1982}. That value assumes the accretion disk is thin and
that hydrostatic equilibrium results from gravitational acceleration only. 
To make sure our model is free of instability \citep{osaki1996} we 
adopt a model
for which those annuli whose standard model $T_{\rm eff}$ values fall below 6300K are reset to 6300K. 
We have	correspondingly
revised the parameters of the warm spot region; the new rim model parameters are in Table~6. For an outer 
annulus temperature
of 6300K, equations (1), (2) lead to a corresponding rim temperature of 5624K outside the warm spot region. 
Irradiation of the rim by the
secondary star raises the rim temperature at the line of (stellar) centers intersection to 5628K.
The temperature profile for the light curve model and the corresponding standard model are in Table~7. 

The `transition temperature' between the hot and cool states falls within a range of values \citep{smak1982}. 
Our light curve model avoids the instability issue by adjusting the temperature of the
outer accretion disk to be at the low end transition temperature, 6300K. 
Our rim
model obviously is a somewhat rough approximation; in view of the uncertainties associated with the vertical
extent of the rim we believe the approximation is appropriate.

Figure~8 shows the comparison of the new synthetic light curve and the photometric observations. The fit is
nearly indistinguishable from that of Figure~5. Although the light curve fits do not distinguish between the
standard model and the model with warmer outer annuli, there are theoretical reasons favoring the latter
model. The standard model does not account for expected warming of the outer accretion disk region when the
transfer stream shock is hidden at a large optical depth \citep{p1977}. The standard model also does not
account for tidal dissipation in the outer accretion disk \citep{lasota2001}. 

We now perform a new simulation of the observed spectra including an allowance for the effect of photospheric temperature
on the WD radius.
Based on L94, we adopted a $T_{\rm eff}$
of 60,000K for the WD. Using \citet[][Figure~4a]{panei2000}, we  determined a radius of
$1.50{\times}10^{-2}R_{\odot}$ for the 60,000K WD.
The WD contribution to the system synthetic spectrum is larger than with the zero temperature WD. We calculated a
new BINSYN accretion disk model based on the revised WD radius. This model required a 29,000K isothermal region,
as compared with the 30,000K isothermal region for the zero temperature WD, for an optimum fit to the observed spectra.
A tabulation of the new model temperature profile and the corresponding
standard model, is in Table~8. Compare with Table~7. The new model has a slightly different profile with the entire
accretion disk on the hot branch. Note that the outer annulus temperature is identical with the
corresponding adopted temperature in Table~7. The new model represents our final model.
Parameters of the final model are in Table~9.

\subsection{Confrontation of final model and observational data}

Figure~9 compares the system synthetic spectrum with the combined $STIS$ and $FUSE$ spectra. The contribution of
a 60,000K WD is at the bottom. The $STIS$ spectrum is the heavier gray plot; the $FUSE$ spectrum is the
lighter one. Except near Ly$\alpha$ and the short wavelength end of the $FUSE$ spectrum the fit is very good, 
allowing for the unmodeled wind lines. 
Based on the line wings, Ly$\alpha$ appears to be in emission, with a central reversal and a superposed
narrow wind line. The strong N V wind line at 1237\AA~depresses what otherwise apparently would be a redward
emission wing of Ly$\alpha$, symmetric with the blueward emission wing.
Figure~10 shows a detail of Figure~9. 

SYNSPEC optionally permits production of a continuum spectrum plus lines of H I and He II. The importance of line
opacity of other atomic species is illustrated by Figure~11, which overplots the continuum spectrum for the
accretion disk plus rim plus line spectrum of the WD on the same observed spectra as Figure~9, and with the same
scaling factor.  

We stress that the divisor, $8.8299{\times}10^{40}$, applied to the synthetic spectrum in placing it in Figure~9
is the appropriate scaling factor if the Hipparcos distance to IX Vel is accurate. It was not determined as an
optimum divisor to produce an empirical fit to the observed spectra; the scaled synthetic spectrum represents 
an absolute flux calibration
for comparison with the observed, reddening corrected, spectra.
The reddening correction,
$E(B-V)=0.01$, was assigned prior to and independently of the Figure~9 plot. 

Different divisors appropriate
to the $\pm~1~\sigma$ uncertainty in the Hipparcos parallax produce synthetic spectrum offsets from the
observed spectra that are approximately 1 dex larger than the Figure~9 residuals. The very good synthetic spectrum
fit was achieved by first adjusting the mass transfer rate and finally by modifying the accretion disk
temperature profile.
The temperatures at the first five division radii, Table~8, are critically important.
(The flux contribution for a given annulus specified within BINSYN is calculated using the temperature at
its inner edge, as given in Table~8.)
Higher temperatures, appropriate to the standard model for the adopted mass transfer rate, not only make the
accretion disk too luminous and too blue, they raise the continuum shortward of Ly~$\alpha$ and appreciably reduce the model
depths of the Lyman series absorption lines. We caution that, since IX Vel varies by up to 0.9 $m_{pg}$ \citep{war1984},
a suitable model at a different epoch could require a different $\dot{M}$ and a different accretion disk temperature
profile.

Figure~12 presents the synthetic continuum spectrum flux over a wide wavelength range, together with the
contributions of the separate objects in the system. The triangles represent flux equivalents of the UBVRI photometry
by \citet{gar1984}; the squares represent flux equivalents of JHKL photometry by \citet{war1984} and \citet{haug1988}. 
Figure~13 presents a detail of Figure~12, showing a portion of the Balmer series. The gray line is an optical
spectrum by \citet{war1983}. We used the curve-following feature of the ADS Dexter facility to extract the plot
from the original publication. Note that the observed spectrum shows narrow emission of H$\beta$ and H$\gamma$
superposed on shallow absorption features.
In producing the
flux equivalents we have used the flux calibrations in Table~1 of \citet{hol2006} for the $U,B,V,R,I,J,H,K$ bands 
and the table on p.202 of
\citet{allen1973} for the L band. 

Figure~14 shows a line spectrum corresponding to Figure~13. For clarity, all synthetic spectra are continuous lines.
As is true in the cases of Figure~9 and Figure~11, the line spectrum falls below the continuum spectrum and, 
in the present instance,
the Figure~13 and Figure~14 synthetic spectra bracket the observed data. 

The secondary star spectrum contributes strongly to the system synthetic spectrum longward of 3000\AA.
It could be argued that a higher $T_{\rm eff}$ for the secondary star would improve the fit. We have used
a coarse representation of the secondary star: a single 3500K log~$g$=4.0 synthetic spectrum calculated with
SYNSPEC from a Kurucz main sequence model. A more accurate secondary star representation would require a much more
sophisticated synthetic spectrum, calculated from a grid of accurate source synthetic spectra covering appropriate
ranges of $T_{\rm eff}$ and log~$g$ and properly simulating the variation of both $T_{\rm eff}$ and log~$g$
over the secondary star photosphere. \citet{how2000} caution that secondary stars may be poorly represented
by main sequence models. Given the uncertainties in the BT parameters we consider the model fit to the
observed data, Figure~12, to be entirely satisfactory.
   
In view of the dependence of synthetic spectrum fit quality on the adopted E(B-V) studied by L94,
we have plotted our final model against the IX Vel observed spectra for E(B-V)=0.0 
({\it i.e.}, the actual observed spectra without a reddening correction), for comparison
with our adopted E(B-V)=0.01. In that case our final model
flux is too large; we produced an optimum fit with a normalizing divisor of $1.00{\times}10^{41}$,
corresponding to an IX Vel distance of 91 pc, slightly closer than one standard deviation from
the Hipparcos distance.
The fit quality is close to that of Figure~9. 
We also plotted our final model against the IX Vel observed
spectra for E(B-V)=0.04. The optimum fit required a normalizing divisor of $6.10{\times}10^{40}$,
corresponding to a distance of 112 pc, slightly farther than one standard deviation beyond the
Hipparcos distance. The synthetic spectrum had a spectral gradient that was
too small.	We believe these comparisons support
our choice of E(B-V)=0.01(\S2).

\section{Discussion}

\subsection{Constraints on ${\dot M}$ from orbital inclination and accretion disk $T_{\rm eff}$ profile}

The geometric constraints from a fit to the K band photometry are of particular importance.
The initial constraint on orbital inclination, $60{\pm}5\arcdeg$, determined from emission line
light curves of H$\alpha$ and H$\beta$ (BT), limits permissible values of $\dot{M}$
by comparison with the $FUSE$ and $STIS$ spectra because of the dependence of system flux on
orbital inclination. In particular, a value
${\dot M}=8{\times}10^{-9}M_{\odot}/{\rm yr}^{-1}$	clearly is too large, and a value 
${\dot M}=5{\times}10^{-9}M_{\odot}/{\rm yr}^{-1}$ is at the lower limit of associated outer boundary
temperature, on the standard model, to avoid accretion disk
instability \citep{smak1982,osaki1996}. Adopting the latter $\dot{M}$, a fit to the
observed spectra requires modification of the standard model temperature profile; 
a satisfactory fit can only be achieved by adopting an isothermal inner portion of the accretion disk.
 
Figure~15 illustrates this point; the plot shows the standard model (unmodified) compared with the
two observed spectra (compare with Figure~9, representing the modified standard model). 
The normalizing divisor is $1.30{\times}10^{41}$; it
places IX Vel more than two
standard deviations farther away than the Hipparcos distance. 
The synthetic spectrum
Lyman lines are much shallower than in
Figure~9. This feature is a result of the very shallow Lyman lines in the higher temperature annulus
spectra and is a separate argument supporting our isothermal region substitution for the standard model
temperature profile. 
The standard model synthetic spectrum 
passes through the tops of the observed spectral
lines shortward of 975\AA, but does not represent the deep Lyman absorption lines in the observed spectra.  

It
is of interest that our isothermal region result agrees with tomography of three of six CV systems studied by
\citet{rut1992} (SW Sex, LX Ser, V1315 Aql), and with analysis results for SDSSJ0809 
\citep{linnell2007}. We are unaware of a theoretical explanation for the departure of the inner annulus
$T_{\rm eff}$ values from the standard model.

It is possible to start with an assumed 
${\dot M}=8{\times}10^{-9}M_{\odot}/{\rm yr}^{-1}$	and, by suitably modifying the adopted
accretion disk temperature profile, produce an acceptable fit to the observed spectra. 
But
the theoretical connection between the adopted temperature profile and the required energy dissipation
budget, to satisfy conservation of energy, becomes more tenuous the larger the departure of the
adopted temperature profile from the standard model. The finally adopted mass transfer rate
preserves as close a connection to the standard model as possible. 

We have verified the unacceptability
of the ${\dot M}=8{\times}10^{-9}M_{\odot}/{\rm yr}^{-1}$ model by converting the temperatures of
the inner five boundaries to 29,000K, the same isothermal value as our adopted model. 
Since the standard model $T_{\rm eff}$ varies
over a finite width annulus, the Table~8 entries list the $T_{\rm eff}$ values at the specific radii
listed in column~1. 
The temperature at the next outer boundary
is 25,871K, so the temperature profile is monotonic.
The corresponding synthetic spectrum is a fairly close match to the observed
spectral gradient, but the flux still is too high. 
Changing the normalizing divisor to $1.40{\times}10^{41}$ produces a fit closely similar to Figure~9.
However, this normalizing divisor corresponds to a distance to IX Vel of 121 parsecs, slightly
more than two standard deviations larger than the Hipparcos distance.

\subsection{An alternative model based on evaporation}

The referee points out a possible alternative model which depends on
inner disk evaporation by the hot WD \citep{lasota2001}. This model would explain why the
standard model produces too much blue flux and also why there appears to be no boundary
layer in the system.

To evaluate this possibility we started with the standard model and the Table~8 $\dot{M}$ and set
successively larger inner truncation radii until we found one for which the system synthetic
spectrum closely matches the de-reddened (E(B-V)=0.01) observed spectra. An inner truncation radius of
$2.335r_{\rm wd}$ produced a system synthetic spectrum with a fit very similar to Figure~9.
The next task is to determine whether the 60,000K WD can evaporate the accretion disk to
that truncation radius. 
 
Evaporation has been discussed in the context of
the quiescent state \citep{mm94,hld99}. This context, with its very low mass transfer rate, 
is not applicable to IX Vel, whose entire accretion disk is on the hot branch. 
Other investigations have been in the context of X-ray binaries
or black hole systems \citep{es97,l1999, ma2000} and \citet{me2000}.
The latter studies provide an evaporation algorithm that permits an estimate of
evaporation effects.
 
The detailed physics of evaporation is poorly understood \citep{lasota2001} 
and it is common to use an
ad hoc prescription, such as the expression used by \citet{me2000} and \citet{dubus2001}:
\begin{displaymath}
{\dot{M}}_{\rm evap}(r)=\frac{0.08{\dot{M}_{\rm Edd}}}{{(r/r_s)^{1/4}}+{\cal E}(r/800r_s)^2},
\end{displaymath} 
where ${\dot{M}_{\rm Edd}}=L_{\rm Edd}/0.1c^2=1.39{\times}10^{18}M_{\rm wd}/M_{\odot}~{\rm g~s^{-1}}$
is the Eddington rate corresponding to the Eddington luminosity, with a 10\% accretion efficiency. 
$c$ is the speed of light and $r_s=2GM_{\rm wd}/c^2$ is the Schwarzschild radius. ${\cal E}$ is an
`evaporation efficiency' factor, with typical values in the range 20 or 30 \citep{me2000} when
applied to the case of soft X-ray transients. 
In the present application the evaporation rate
should equal the mass transfer rate from the secondary, evaluated at the truncation radius.
The mass transfer rate from the secondary star is $3.2{\times}10^{17}~{\rm gm}~{\rm s}^{-1}$.
For ${\cal E}=20.0$, ${\dot{M}}_{\rm evap}=2.2{\times}10^{13}~{\rm g~s^{-1}}$ at
$r/r_{\rm wd}=2.335$. 
Based on this model, evaporation is four 
orders of magnitude too
small to produce the required disk truncation.

\subsection{System wind}

IX Vel shows prominent effects of a wind associated with the system
\citep[][and L94]{sion1985,mauche1991,prinja1995,hartley2002}.
\citet{hartley2002} study whether the IX Vel wind is line-driven. They find that the strong
correlation between UV brightness and wind activity predicted by line-driven wind models 
\citep{pereyra2006} is
disobeyed for both IX Vel and V3885 Sgr, and conclude that the most promising alternative to a
pure line-driven disk model is a hybrid model including a combination of MHD and line forces.
\citet{proga2003a,proga2003b} calculates resonance-line profiles for radiation-driven winds and
produces a good fit to the observed C IV ${\lambda}1549$ line, with its P Cygni type profile, 
but requires a mass transfer rate
that is six times larger than determined in this study.	Other wind lines are much stronger than
the C IV ${\lambda}1549$ line and do not show P Cygni characteristics. 

\citet{longk2002} describe
an improved code for spectral synthesis of wind lines in the non-MHD case. 
See \citet{murray2002}
for a comparison of
line-driven winds and MHD driving. 

The standard model accounts for 1/2 of the potential energy liberated from the mass transfer
stream \citep{fr92}. A boundary layer often is proposed as the mechanism to account for the
balance (FKR), but we have seen that IX Vel does not have a hot boundary layer. We assume the system wind
carries off the liberated potential energy not accounted for by the accretion disk faces and rim. 

\subsection{The bright spot}

\citet{smak2002}, hereafter S2002, discusses the structure of the outer parts of accretion disks
with particular reference to tidal heating and radiation from the rim, including effects
of a bright spot. S2002 proposes that the luminosity of the rim arises entirely from tidal heating,
and consumes all of that source of energy,
and that a local bright spot radiates away all of the energy released directly by the stream-disk
interaction. S2002 references the tidal heating theory in \citet{lasota2001}, and \citet{bm2001}.
The temperature of the S2002 accretion disk face would be given by the standard model.
S2002 points to the study by \citet{ich1994} which finds that tidal torques are limited to a
radially narrow region, producing a sharply-defined accretion disk edge, which S2002 takes to
be cylindrical. This model differs from the model we use to define the disk edge \citep{hub1991}.
The impact of the mass transfer stream produces strong turbulence and affects a fairly large radial
region of the accretion disk (see below). Consequently, we believe the S2002 model is oversimplified.

S2002 calculates a bright spot temperature in the approximate range 18,000K-20,000K, a range supported
by an analysis of UX UMa \citep{smak1994} for which there is clear evidence for prompt emission of
the energy liberated by the stream-disk collision. This temperature range accords with values determined
for other CV systems \citep[][sect. 2.6.5]{w1995}.
The K band photometry rules out a IX Vel bright spot
temperature as high as the S2002 model and, instead, limits the bright spot temperature to $\approx6100$K,
only about 500K warmer than the rim regions unaffected by the bright spot. (Note that our rim model assigns
a rim temperature, 5624K, somewhat lower than the outermost annulus temperature, 6300K.)

The study by \citet{livio1986} considers the degree of stream penetration
into the rim based on their parameter $\alpha$. They show that the stream produces appreciable turbulence
in the outer annuli, and that, for small $\alpha$, there can be large penetration, leading to energy release
within the outer annuli. Their analysis considers the case in which the stream has a larger diameter than
the rim height.	We show below that, in IX Vel, the stream diameter is less than the rim height.
The ratio of the stream radius, $r_s$, to the component separation, D, near the L1 point, is 
given by the small parameter $\epsilon$ where \citep{cook1984}

\begin{equation}
{\epsilon}{\approx}6.1{\times}10^{-3}\Big{(}\frac{T_{\rm eff}}{10^3}\Big{)}^{1/2}\Big{(}\frac{M_2}{M_0}\Big{)}^{-1/3}
\Big{(}\frac{P_{\rm orb}}{1 {\rm~hr}}\Big{)}^{1/3}(1+q)^{-1/3},	  q=M_2/M_1.
\end{equation}
In this equation $T_{\rm eff}$ applies to the secondary star.
In the case of IX Vel, $\epsilon=0.02$ and $r_s=2.2\times10^9$~cm. From Table~4, the TLUSTY height of the rim
is $5.4\times10^9$~cm, and the adopted height is about two times this. 
Thus, the rim height is larger than the stream cross-section and
conditions differ from those studied by Livio, Soker, \& Dgani. 
\citet{rs1987,rs1988} have developed a hydrodynamical model of the stream-disk interaction.	On their model,
the stream density at a distance $r$ from the symmetry axis is

\begin{equation}
{\rho}_s={\rho}_{sc}~{\rm exp}(-r^2/r_s^2),
\end{equation}
where $\rho_{sc}$ is the stream density on the symmetry axis. \citet{rs1987} set 
$\rho_{sc}=1.7\times10^{-9}~{\rm gm}~{\rm cm}^{-3}$ for $\dot{M}=10^{-9}~M_{\odot}{\rm yr}^{-1}$.
For the IX Vel $\dot{M}$, (five times larger) we have $\rho_{sc}=8.5\times10^{-9}~{\rm gm}~{\rm cm}^{-3}$.
The TLUSTY model for the outermost annulus has a central plane density of $1.47{\times}10^{-6}~{\rm gm}~{\rm cm}^{-3}$
and a density at the top of $4.25\times10^{-14}~{\rm gm}~{\rm cm}^{-3}$. The density at $z=2.2{\times}10^9~{\rm cm}$
(the stream radius) is $5.0{\times}10^{-6}~{\rm gm}~{\rm cm}^{-3}$.
The top of the annulus is 
at a height of $6.32{\times}r_s$.
Using equation~(5), ${\rho}_s=3.49{\times}10^{-26}~{\rm gm}~{\rm cm}^{-3}$ at the rim. Assuming the annulus scale 
height increases by the
same factor as the increased vertical rim dimension, in meeting the geometric requirements to shadow the secondary,
we find that the accretion disk rim is everywhere more dense than the
stream, consistent with both R\'{o}\.{z}yczka \& Schwarzenberg-Czerny and Livio et al. A basic difference of
IX Vel from these two studies is that the rim height is appreciably greater than the stream radius. 
There do not appear to be any studies of this case.

\citet{p1977} argues that,
if the radius of the incident stream is less than the rim semi-height,
the impacting stream plows into the rim and dissipates its energy within the accretion disk.  
An important point made by \citet{p1977}
is that, when the stream penetrates deeply, there is a large optical depth to the region of energy dissipation.
This argument provides a physical basis to explain why IX Vel has a warm spot rather than a higher temperature spot. 
It is clear that, for IX Vel, the stream-disk interaction must produce some heating of the outer accretion disk.

The size of the bright spot region is somewhat extended, compared with prominent spots in other CVs.
For example, \citet{cook1985} found an angular width of the spot in OY Car of up to $30\arcdeg$ and an angle
of about $5\arcdeg$ between the impact point of the stream and the maximum flux from the spot. Compare with Table~6.
Although bright spots in a number of systems associate flickering in those systems
\citep{war1985} with the warm spot \citep[][sect. 2.6.5]{w1995}, there is no basis to connect the flickering
observed in IX Vel \citep{gar1984,eggen1984,war1985} with the warm spot.

It is of interest that \citet{kubiak1999} observe enhanced emission in the Paschen H lines and the CaII
triplet (8498.02\AA, 8542.09\AA, 8662.14\AA) at orbital phase 0.32, nearly diametrically across the disk
from the warm rim region. The enhanced emission parallels other examples of excitation on the opposite
side of the disk from the bright spot location \citep{szkody1992, livio1993, hillier2000}. Our model
would not show significant phase-resolved spectral variation and BINSYN currently does not model emission
lines so we have no basis to simulate that observational result.

\subsection{The annulus extended vertical heights}

It now is generally assumed that the
source of viscous dissipation in accretion disks arises from MRI \citep{bh1991}.
\citet{hirose2006,blaes2006} show that magnetic pressure can be important in vertical hydrostatic support and can
lead to vertically extended annuli by about a factor two compared with analytic models. 
This result is roughly equal to the ratio of our empirically required vertical extent of the outer annulus
to the theoretical height
based on the TLUSTY models.	Since the optical thickness of the annuli is large (see Table~4), 
the predicted spectrum is quite insensitive to the actual optical thickness of the disk.

\subsection{Caveats}

We note the many uncertainties in the physics of accretion disks and the simplifications required
(MHD effects, our adoption of a single viscosity 
$\alpha$ for the entire accretion disk,
our neglect of radial energy transport, our assumption of vertical hydrostatic equilibrium - and consequent 
neglect of wind effects, etc.).
There also are many
approximations in our geometrical model (e.g., neglect of the semi-transparent outer 
layers of the disk
when computing the final spectrum).

The intrinsic variability of the source is an important consideration when deciding how worthwhile it is
to find a precise fit to our particular set of observations.
For example, a discussion within our group raised the question: In view of all the uncertainties listed above,
why not simply use the most appropriate \citet[hereafter WH]{wade1998} accretion disk model, with WD spectrum 
added, to
represent the system? A test with the WH model cc, whose 
${\dot M}=3.2{\times}10^{-9}M_{\odot}/{\rm yr}^{-1}$, and
for an adopted inclination of $60\arcdeg$, 
showed that a fair fit to the observed spectra
could be achieved, with a calculated distance to IX Vel of 89 pc for an added 55,000K WD, only different from
the Hipparcos distance by about one standard deviation. And the cc model is a standard model. So, can we
maintain the objective reality implied by our final model departure from a standard model temperature profile?

The response in this debate was: The cc model cuts off at $r/r_{\rm wd,0}=21.7$ because WH had decided
to terminate annulus models at a $T_{\rm eff}$ of about 10,000K. The flux error in neglecting the outer
accretion disk, as WH note, is of order 9\%. Our calculation, with annuli extending essentially to the tidal cutoff,
is not subject to that error. But the more
important point is that the tidal cutoff radius is at $r/r_{\rm wd,0}=57.1$. Calculation of the full standard model
at the model cc mass transfer rate shows that a substantial outer range of the accretion disk has a $T_{\rm eff}$
below the transition temperature to the low state: the cc model is unstable to flux outbursts, where we have taken
the transition temperature as 6300K. Recent developments \citep{lasota2001,bm2001} show that the standard model
needs correction for tidal and stream impact heating, and these additions affect accretion disk susceptability
to flux outbursts. 
Parenthetically,
the WH annuli were calculated with the viscosity represented by a Reynolds number of 5000; models using the
now generally adopted \citet{ss1973} $\alpha=0.1$ can have a very different structure, and would alter the Table~4
entries, although the $T_{\rm eff}$
values would be the same. Moreover, use of the cc model provides no means to calculate a synthetic light curve, 
as in Figure~8,
and to simulate properties of the accretion disk rim, including its extended vertical height, as our final model
has done. Finally, \citet{rut1992} provide observational evidence supporting departures from the standard model.
If ${\dot M}$ varies, as the larger (i.e., larger than flickering) IX Vel photometric variability implies, the 
accretion disk will have a transient 
response to each change; depending on the transient response time, the accretion disk may well be in a perpetual 
state of departure from the standard model.
It is desirable to have modeling procedures capable of simulating such departures.

The other side in the debate has a valid point:
We have not produced an array of models, covering the error ranges of $M_{\rm wd}$, ${\dot M}$, $i$, the 
Hipparcos distance, etc., in
search of an optimum model, together with confidence ellipses for all of the variables involved. 
It is entirely possible, even likely, that such a search would find other models
that fit the observed spectra as well as our final model. The computational effort to do that is
completely prohibitive.

What we have done, for the data set we have studied,
is to show that a modification of the standard model produces a better fit to the available constraints
(observed spectra, K band light curve, Hipparcos distance) than the standard model, and that our final model
is a very good fit to all of the constraints except for a short FUV wavelength region. 

\section{Summary}

This analysis of IX Vel leads to the following conclusions:

(1) At the times of the $FUSE$ and $STIS$ spectra studied here, our model derives a mass transfer rate 
from the secondary star
of ${\dot M}=5{\times}10^{-9}M_{\odot}/{\rm yr}^{-1}$ and excludes a rate as large as
${\dot M}=8{\times}10^{-9}M_{\odot}/{\rm yr}^{-1}$.

(2) It is possible to achieve a close synthetic spectrum fit to the observed $FUSE$ and $STIS$ spectra
by adopting an accretion disk temperature profile which agrees with the standard model except for the
inner annuli, extending to $r/r_{\rm wd}{\approx}4$. Shortward of that radius the
accretion disk is approximately isothermal. 

(3) A light synthesis fit to K band photometry requires shallow eclipses of the accretion disk rim and 
secondary star limb. The geometry, including the calculated outer radius of the accretion disk, constrains
the orbital inclination to $i=57{\pm}2{\arcdeg}$(est.).

(4) A synthetic light curve agreeing with the observed ellipsoidal (not Algol type) variation,
requires a vertically extended accretion disk rim to shadow the secondary star, beyond that 
predicted by gravitational
equilibrium. The enhanced vertical extension is consistent with recent MHD predictions for CV accretion disks.

(5) Representation of differing heights of alternate K band maxima requires a warm rim region downstream of
the intersection point of the mass transfer stream with the accretion disk rim. The location and temperature
of the warm region are inconsistent with prompt radiation from a bright spot associated with 
a rim shock, and consistent with deep penetration of the stream.

The authors thank the anonymous referee for carefully reading the manuscript and providing a valuable
critique; responding to it has significantly improved the paper presentation.
P.S. acknowledges support from $FUSE$ grant NNG04GC97G and HST grant GO-09724. 
PG is thankful to the Space Telescope Science Institute for its kind
hospitality. 
Support for this work was provided by NASA through grant number 
HST-AR-10657.01-A  to Villanova University (P. Godon) from the Space
Telescope Science Institute, which is operated by the Association of
Universities for Research in Astronomy, Incorporated, under NASA
contact NAS5-26555. This research was partly based on observations made with
the NASA-CNES-CSA Far Ultraviolet Spectroscopic Explorer. FUSE is operated
for NASA by the Johns Hopkins University under NASA contract NAS5-32958.

%The vertical extent of the rim is in agreement with the calculated extra extension on allowing
%for magnetic support.
%
%The light curve analysis confirms the theoretical tidal truncation radius.
%
%The fit to the observed spectra confirms the lower temperatures of the inner annuli.

\clearpage


%%%%%%%%%%%%%%%%%%%%%%%%%%%%%%%%%%%%%%%%%%%%%%%%%%%%%%%%%%%%%%%%%%%
%%% TABLES
%%%%%%%%%%%%%%%%%%%%%%%%%%%%%%%%%%%%%%%%%%%%%%%%%%%%%%%%%%%%%%%%%%%

%%%%%%%%%%%%%%%%%%%%%%%%%%%%%%%%%%%%%%%%%%%%%%%%%%%%%%%%%%%%%%%%%%%

\begin{table} 
\caption{FUV Observations of IX Vel: HST \& FUSE Spectra} 
\begin{tabular}{clcllll}
\hline
 Date      & Telescope   & Exp time & Dataset   & Filter/Grating &Operation & Calibration  \\ 
(dd/mm/yyyy) & /Instrument & (sec)  &           & /Aperture      & Mode     & Software \\ 
\hline
 15-04-2000 & FUSE        & 7336     & Q1120101  & LWRS           & HIST     & CalFUSE 3.0.7 \\ 
 19-07-2000 & STIS        & 1750     & o5bi03010 & E140M/0.1x0.03 & TIME-TAG & CALSTIS 2.22 \\  
\hline
\end{tabular}
\end{table}

%%%%%%%%%%%%%%%%%%%%%%%%%%%%%%%%%%%%%%%%%%%%%%%%%%%%%%%%%%%%%%%%%%%

\clearpage

%%%%%%%%%%%%%%%%%%%%%%%%%%%%%%%%%%%%%%%%%%%%%%%%%%%%%%%%%%%%%%%%%%%
\begin{table} 
\caption{FUSE Lines} 
\begin{tabular}{lll}
\hline
Line           & Wavelength   & Origin   \\
Identification & (\AA )       &           \\
\hline
H\,{\sc i}    & see text      & s,c,ism   \\
O\,{\sc i}    & see text      & a,c,ism   \\
N\,{\sc iv}   & 921.5-924.9   &    s      \\
S\,{\sc vi}   & 933.5         &    s      \\
S\,{\sc vi}   & 944.5         &    s      \\
N\,{\sc i}    & 952.4-954.1   & a,c,ism  \\
              & 964.0-965.04  &  c,ism  \\
C\,{\sc iii}  & 977.0         &    s      \\
N\,{\sc iii}  & 991.6         &    s      \\
O\,{\sc vi}   & 1031.9        &   s       \\
              & 1037.6        &   s       \\
C\,{\sc ii}   & 1036.3        &   ism     \\
Ar\,{\sc i}   & 1048.2        &   ism     \\ 
              & 1066.7        &  ism     \\ 
S\,{\sc iv}   & 1062.6        &    s      \\
S\,{\sc iv}   & 1073.0        &    s      \\ 
              & 1073.5        &    s      \\
N\,{\sc ii}   & 1084.0        &  a,c,ism   \\
S\,{\sc iv, vi}& 1117.8        &   s      \\ 
P\,{\sc v}    & 1118.0        &   s    \\ 
              & 1128.0 &   s      \\ 
Si\,{\sc iv}  & 1122.5 & s \\ 
              & 1128.3 &   s       \\
N\,{\sc i}    & 1134.2-1135.0  & a,c,ism   \\
Fe\,{\sc ii}  & 1144.9        &  c,ism    \\
C\,{\sc iii}  & 1174.9-1176.4 &   s       \\
\hline
\end{tabular}
\tablenotetext{}{a=possibly with terrestrial contamination; 
c=circumbinary material; ism=ISM; s=source   
}
\end{table}

%%%%%%%%%%%%%%%%%%%%%%%%%%%%%%%%%%%%%%%%%%%%%%%%%%%%%%%%%%%%%%%%%%%
\clearpage
%%%%%%%%%%%%%%%%%%%%%%%%%%%%%%%%%%%%%%%%%%%%%%%%%%%%%%%%%%%%%%%%%%%

\begin{table} 
\caption{STIS Lines} 
\begin{tabular}{lll}
\hline
Line           & Wavelength   & Origin     \\
Identification & (\AA )       &            \\
\hline
C\,{\sc iii}  & 1174.9-1176.4 &   s        \\
Si\,{\sc ii}  & 1190.4        &  c,ism     \\
              & 1193.3        &  c,ism     \\
              & 1194.5        &  c,ism     \\
N\,{\sc i}    & 1199.5-1200.7 &  c,ism   \\  
Si\,{\sc iii} & 1206.5        &    c       \\
H\,{\sc i}    & 1215.7        &    s       \\
N\,{\sc v}    & 1238.8 & s \\ 
              & 1242.8 &   s       \\
S\,{\sc ii}   & 1250.6 & c,ism \\ 
              & 1253.8 & c,ism     \\ 
              & 1259.5 & c,ism \\ 
              & 1260.4 & c,ism     \\ 
0\,{\sc i}    & 1302.2 & c,ism   \\ 
              & 1304.9 & c,ism    \\
              & 1306.0 & c,ism    \\
C\,{\sc ii}   & 1334.5 & c,ism    \\ 
              & 1335.7 & c,ism     \\
Si\,{\sc iv}  & 1393.8        &   s        \\
              & 1402.8        &   s        \\
C\,{\sc iii}  & 1426.4  & c,ism        \\
Si\,{\sc ii}  & 1526.7        &   c,ism    \\
              & 1533.4        &   c,ism    \\
C\,{\sc iv}   & 1548.2        &   s        \\
              & 1550.8        &   s        \\
He\,{\sc ii}  & 1640.5        &   s        \\
Al\,{\sc ii}  & 1670.8        &   c,ism    \\
\hline
\end{tabular}
\tablenotetext{}{c=circumbinary material; ism=ISM; s=source   
}
\end{table}

%%%%%%%%%%%%%%%%%%%%%%%%%%%%%%%%%%%%%%%%%%%%%%%%%%%%%%%%%%%%%%%%%%%
\clearpage
%%%%%%%%%%%%%%%%%%%%%%%%%%%%%%%%%%%%%%%%%%%%%%%%%%%%%%%%%%%%%%%%%%%

\begin{deluxetable}{rrrrrrrr}
\tablewidth{0pt}
\tablenum{4}
\tablecaption{Properties of accretion disk with mass transfer rate 
$\dot{M}=5.0{\times}10^{-9}~{M}_{\odot}{\rm yr}^{-1}$ and WD mass of $0.80{M}_{\odot}$.}
\tablehead{	  
\colhead{$r/r_{\rm wd,0}$} & \colhead{$T_{\rm eff}$} & \colhead{$m_0$} 
& \colhead{$T_0$} & \colhead{log~$g$}
& \colhead{$z_0$} & \colhead{$Ne$} & \colhead{{$\tau_{\rm Ross}$}}}
\startdata
1.36  &  59317  &  1.213E4   &  39973  &  6.97   & 7.81E7  & 9.68E11  & 1.99E4\\
2.00  &	 53184  &  1.481E4	 &  37066  &  6.71   & 1.35E8  & 5.68E11  & 2.37E4\\
3.00  &	 43006  &  1.447E4	 &  31573  &  6.41   & 2.29E8  & 3.35E11  & 2.80E4\\
4.00  &	 36147  &  1.347E4	 &  24987  &  6.19   & 3.25E8  & 2.47E11  & 3.23E4\\
5.00  &	 31354  &  1.251E4	 &  19127  &  6.02   & 4.24E8  & 2.13E11  & 3.67E4\\
6.00  &	 27816  &  1.169E4	 &  15966  &  5.87   & 5.28E8  & 1.84E11  & 4.15E4\\
8.00  &	 22919  &  1.037E4	 &  13021  &  5.65   & 7.47E8  & 1.34E11  & 5.17E4\\
10.00 &	 19661  &  9.380E3	 &  11433  &  5.48   & 9.76E8  & 1.01E11  & 6.28E4\\
12.00 &	 17318  &  8.610E3	 &  10386  &  5.34   & 1.21E9  & 7.80E10  & 7.42E4\\
14.00 &	 15542  &  7.990E3	 &   9599  &  5.22   & 1.46E9  & 6.34E10  & 8.53E4\\
16.00 &	 14143  &  7.479E3	 &   8970  &  5.11   & 1.70E9  & 5.31E10  & 9.50E4\\
18.00 &	 13009  &  7.048E3	 &   8448  &  5.02   & 1.95E9  & 4.53E10  & 1.02E5\\
20.00 &	 12068  &  6.679E3	 &   8000  &  4.93   & 2.19E9  & 3.91E10  & 1.06E5\\
24.00 &	 10591  &  6.076E3	 &   8594  &  4.77   & 2.64E9  & 2.60E10  & 1.06E5\\
30.00 &	  9019  &  5.401E3	 &   6478  &  4.59   & 3.37E9  & 1.74E10  & 9.85E4\\
35.00 &	  8067  &  4.973E3	 &   6546  &  4.46   & 3.82E9  & 1.37E10  & 9.37E4\\
40.00 &	  7322  &  4.626E3	 &   5945  &  4.42   & 4.35E9  & 6.55E09  & 9.14E4\\
45.00 &	  6721  &  4.338E3	 &   5470  &  4.39   & 4.90E9  & 2.26E09  & 9.40E4\\
50.00 &   6224  &  4.094E3   &   5067  &  4.13   & 5.41E9  & 6.81E08  & 1.05E5\\
\enddata
\tablecomments{Each line in the table represents a separate annulus.
A \citet{ss1973} viscosity parameter $\alpha=0.1$ was used in calculating all annuli.
The WD radius, $r_{\rm wd,0}$, is the radius of a zero temperature Hamada-Salpeter carbon model.
}		 
\end{deluxetable}

%%%%%%%%%%%%%%%%%%%%%%%%%%%%%%%%%%%%%%%%%%%%%%%%%%%%%%%%%%%%%%%%%%%

\clearpage

%%%%%%%%%%%%%%%%%%%%%%%%%%%%%%%%%%%%%%%%%%%%%%%%%%%%%%%%%%%%%%%%%%%
\begin{deluxetable}{llll}
\tablewidth{0pt}
\tablenum{5}
\tablecaption{IX Vel Model System Parameters, \citet{beu1990} }
\tablehead{
\colhead{parameter} & \colhead{value} & \colhead{parameter} & \colhead{value}}
\startdata
$M_{\rm wd}$  &  $0.80{\pm}0.15M_{\odot}$	 & $i$    &  $60{\pm}5{\arcdeg}$\\
${M}_2$  &  $0.52{\pm}0.10{M}_{\odot}$	 & $K_{\rm wd}$	&	138~${\rm km}~{\rm s}^{-1}$\\
${\dot{M}}$      &  $7.9{\pm}1.0{\times}10^{-9}{M}_{\odot} {\rm yr}^{-1}$ &	 $K_2$	&	212~${\rm km}~{\rm s}^{-1}$\\
P    &  0.1939292 days	  &  	 $T_{\rm eff,2}$	&	3500K\\
$d$              &  $95{\pm}12$~pc\\	 
\enddata
%\tablecomments{${\rm wd}$ refers to the WD; $s$ refers to the secondary star.
\end{deluxetable}

%%%%%%%%%%%%%%%%%%%%%%%%%%%%%%%%%%%%%%%%%%%%%%%%%%%%%%%%%%%%%%%%%%%
\clearpage
%%%%%%%%%%%%%%%%%%%%%%%%%%%%%%%%%%%%%%%%%%%%%%%%%%%%%%%%%%%%%%%%%%%

\begin{deluxetable}{lrlrlr}
\tablewidth{0pt}
\tablenum{6}
\tablecaption{Properties of accretion disk rim model, K band light curve 
}
\tablehead{	  
\colhead{parameter} & \colhead{value} & \colhead{parameter} & \colhead{value}
& \colhead{parameter} & \colhead{value}
} 
\startdata
$T_{\rm eff,rim}$  &  5624K	&	HSAZ	&	270 	&	HWDW	&	30\\
HSTEMP  &	 6.1K	&	HSDWD	&	10	&	TDWND	&	5.7K\\
HSUP  &	  10	&	TUPD	&	5.7K\\
\enddata
\tablecomments{
HSAZ is the position angle (degrees) of the center of the bright spot region.
HWDW is the angular width of the bright spot, in degrees.
HSTEMP is the temperature of the bright spot.
HSDWD is the angular width (degrees) of the downwind region from the end of the bright spot
to the merge point with the rim proper.
TDWND is the temperature of the middle of region HSDWD.
HSUP is the angular width (degrees) of the upwind region comparable to HSDWD.
TUPD is the temperature of the middle of region HSUP.
The value of $T_{\rm eff,rim}$ corresponds, according to equations (1), (2) to an outer
annulus $T_{\rm eff}$ of 6300K.
}		 
\end{deluxetable}

%%%%%%%%%%%%%%%%%%%%%%%%%%%%%%%%%%%%%%%%%%%%%%%%%%%%%%%%%%%%%%%%%%%
\clearpage
%%%%%%%%%%%%%%%%%%%%%%%%%%%%%%%%%%%%%%%%%%%%%%%%%%%%%%%%%%%%%%%%%%%
\begin{deluxetable}{rrrrrr}
\tablewidth{0pt}
\tablenum{7}
\tablecaption{Comparison of temperature profiles, adopted and standard model, for 
accretion disk with mass transfer rate of
$\dot{M}=5.0{\times}10^{-9}~{M}_{\odot}{\rm yr}^{-1}$ and WD mass of $0.80{M}_{\odot}$.}
\tablehead{	  
\colhead{$r/r_{\rm wd,0}$} & \colhead{$T_{\rm eff}$(adopt, K)} & \colhead{$T_{\rm eff}$(std, K)} 
& \colhead{$r/r_{\rm wd,0}$} & \colhead{$T_{\rm eff}$(adopt, K)}
& \colhead{$T_{\rm eff}$(std, K)}}
\startdata
1.00    &	30000	&	54961	&	29.23	&	9206	&	9206\\
1.18	&	30000	&	55252	&	31.08	&	8805	&	8805\\
1.36	&	30000	&	59557	&	32.94	&	8443	&	8443\\
3.22	&	30000	&	41366	&	34.80	&	8114	&	8114\\
5.08	&	30000	&	31111	&	36.66	&	7814	&	7814\\
6.93	&	25301	&	25301	&	38.51	&	7639	&	7639\\
8.79	&	21527   &	21527	&	40.37	&	7285	&	7285\\
10.65	&	18856	&	18856	&	42.23	&	7051	&	7051\\
12.51	&	16854	&	16854	&	44.09	&	6833	&	6833\\
14.36	&	15291	&	15291	&	45.94	&	6631	&	6631\\
16.22	&	14030	&	14030	&	47.80	&	6442	&	6442\\
18.08	&	12991	&	12991	&	49.66	&	6300	&	6266\\
19.94	&	12116	&	12116	&	51.52	&	6300	&	6100\\
21.80	&	11368	&	11368	&	53.38	&	6300	&	5944\\
23.65	&	10720	&	10720	&	55.23	&	6300	&	5797\\
25.51	&	10154	&	10154	&	57.09	&	6300	&	5659\\
27.37	&	9653	&	9653\\
\enddata
\tablecomments{
The WD radius, $r_{\rm wd,0}$, is the radius of a zero temperature Hamada-Salpeter carbon model.
}		 
\end{deluxetable}

%%%%%%%%%%%%%%%%%%%%%%%%%%%%%%%%%%%%%%%%%%%%%%%%%%%%%%%%%%%%%%%%%%%
\clearpage
%%%%%%%%%%%%%%%%%%%%%%%%%%%%%%%%%%%%%%%%%%%%%%%%%%%%%%%%%%%%%%%%%%%
\begin{deluxetable}{rrrrrr}
\tablewidth{0pt}
\tablenum{8}
\tablecaption{Comparison of temperature profiles, adopted and standard model, for 
accretion disk with mass transfer rate of
$\dot{M}=5.0{\times}10^{-9}~{M}_{\odot}{\rm yr}^{-1}$ and WD mass of $0.80{M}_{\odot}$.}
\tablehead{	  
\colhead{$r/r_{\rm wd}$} & \colhead{$T_{\rm eff}$(adopt, K)} & \colhead{$T_{\rm eff}$(std, K)} 
& \colhead{$r/r_{\rm wd}$} & \colhead{$T_{\rm eff}$(adopt, K)}
& \colhead{$T_{\rm eff}$(std, K)}}
\startdata
1.00    &	29000	&	41632	&	19.36	&	9015	&	9015\\
1.18	&	29000	&	41632	&	20.56	&	8636	&	8636\\
1.36	&	29000	&	43303	&	21.76	&	8293	&	8293\\
2.56	&	29000	&	34310	&	22.96	&	7980	&	7980\\
3.76	&	27415	&	27415	&	24.16	&	7694	&	7694\\
4.96	&	23003	&	23003	&	25.36	&	7430	&	7430\\
6.16	&	19950   &	19950	&	26.56	&	7187	&	7187\\
7.36	&	17703	&	17703	&	27.77	&	6962	&	6962\\
8.56	&	15972	&	15972	&	28.97	&	6753	&	6753\\
9.76	&	14594	&	14594	&	30.17	&	6558	&	6558\\
10.96	&	13466	&	13466	&	31.37	&	6376	&	6376\\
12.16	&	12524	&	12524	&	32.57	&	6300	&	6204\\
13.36	&	11724	&	11724	&	33.77	&	6300	&	6044\\
14.56	&	11035	&	11035	&	34.97	&	6300	&	5893\\
15.76	&	10434	&	10434	&	36.17	&	6300	&	5751\\
16.96	&	 9906	&	 9906	&	37.37	&	6300	&	5616\\
18.16	&	 9436	&	 9436\\
\enddata
\tablecomments{$r_{\rm wd}$ in the first column heading represents the radius
of a 60,000K carbon WD.}

\end{deluxetable}

%%%%%%%%%%%%%%%%%%%%%%%%%%%%%%%%%%%%%%%%%%%%%%%%%%%%%%%%%%%%%%%%%%%
\clearpage
%%%%%%%%%%%%%%%%%%%%%%%%%%%%%%%%%%%%%%%%%%%%%%%%%%%%%%%%%%%%%%%%%%%
\begin{deluxetable}{llll}
\tablewidth{0pt}
\tablenum{9}
\tablecaption{IX Vel Model System Parameters}
\tablehead{
\colhead{parameter} & \colhead{value} & \colhead{parameter} & \colhead{value}}
\startdata
${ M}_{\rm wd}$  &  $0.80{\pm}0.2{M}_{\odot}$	 & $T_{\rm eff,s}$(pole)    &  $3500{\pm}1000$K \\
${M}_{\rm sec}$  &  $0.52{\pm}0.10{M}_{\odot}$	 & $T_{\rm eff,s}$(point)	&	3119K\\
${\dot{M}}$      &  $5.0{\pm}1.0{\times}10^{-9}{M}_{\odot} {\rm yr}^{-1}$ &	 $T_{\rm eff,s}$(side)	&	3451K\\
P    &  0.1939292 days	  &  	 $T_{\rm eff,s}$(back)	&	3345K \\
$D$              &  $1.54588R_{\odot}$ & $r_s$(pole) &  $0.496R_{\odot}$\\	 
${\Omega}_{\rm wd}$         & 103.8 & $r_s$(point)   & $0.705R_{\odot}$\\
${\Omega}_s$                &  3.15422 & $r_s$(side)  & $0.3518R_{\odot}$\\
{\it i}              &   $57{\pm}2{\degr}$& $r_s$back)   & $0.568R_{\odot}$\\
$T_{\rm eff,wd}$         &  $60,000{\pm}10000$K  & log $g_s$(pole) & 4.78\\
$r_{\rm wd}$      &   $0.0150R_{\odot}$ & log $g_s$(point) & -4.63\\
log $g_{\rm wd}$  &   7.99& log $g_s$(side)  & 4.70\\
$A_{\rm wd}$        &  1.0 & log $g_s$(back)  & 4.54\\ 
$A_s$               &  0.6  & $r_a$ & $0.56R_{\odot}$\\
${\beta}_{\rm wd}$  &  0.25 & $r_b$ & $0.00985R_{\odot}$\\
${\beta}_s$         &  0.08& $H$    & $0.20R_{\odot}$\\
\enddata
\tablecomments{${\rm wd}$ refers to the WD; $s$ refers to the secondary star.
$D$ is the component separation of centers,
${\Omega}$ is a Roche potential. Temperatures are polar values, 
$A$ values are bolometric albedos, and $\beta$ values are 
gravity-darkening exponents. 
$r_a$ specifies the outer radius 
of the accretion disk, set at the tidal cut-off radius, 
and $r_b$ is the accretion disk inner radius, as 
determined in the final system model, while $H$ is 
the semi-height of the accretion disk rim.}  
\end{deluxetable}

%%%%%%%%%%%%%%%%%%%%%%%%%%%%%%%%%%%%%%%%%%%%%%%%%%%%%%%%%%%%%%%%%%%

%%%%%%%%%%%%%%%%%%%%%%%%%%%%%%%%%%%%%%%%%%%%%%%%%%%%%%%%%%%%%%%%%%%
%%% FIGURES
%%%%%%%%%%%%%%%%%%%%%%%%%%%%%%%%%%%%%%%%%%%%%%%%%%%%%%%%%%%%%%%%%%%

\clearpage

%%%%%%%%%%%%%%%%%%%%%%%%%%%%%%%%%%%%%%%%%%%%%%%%%%%%%%%%%%%%%%%%%%%
\begin{figure}[tb]
\epsscale{0.97}
\plotone{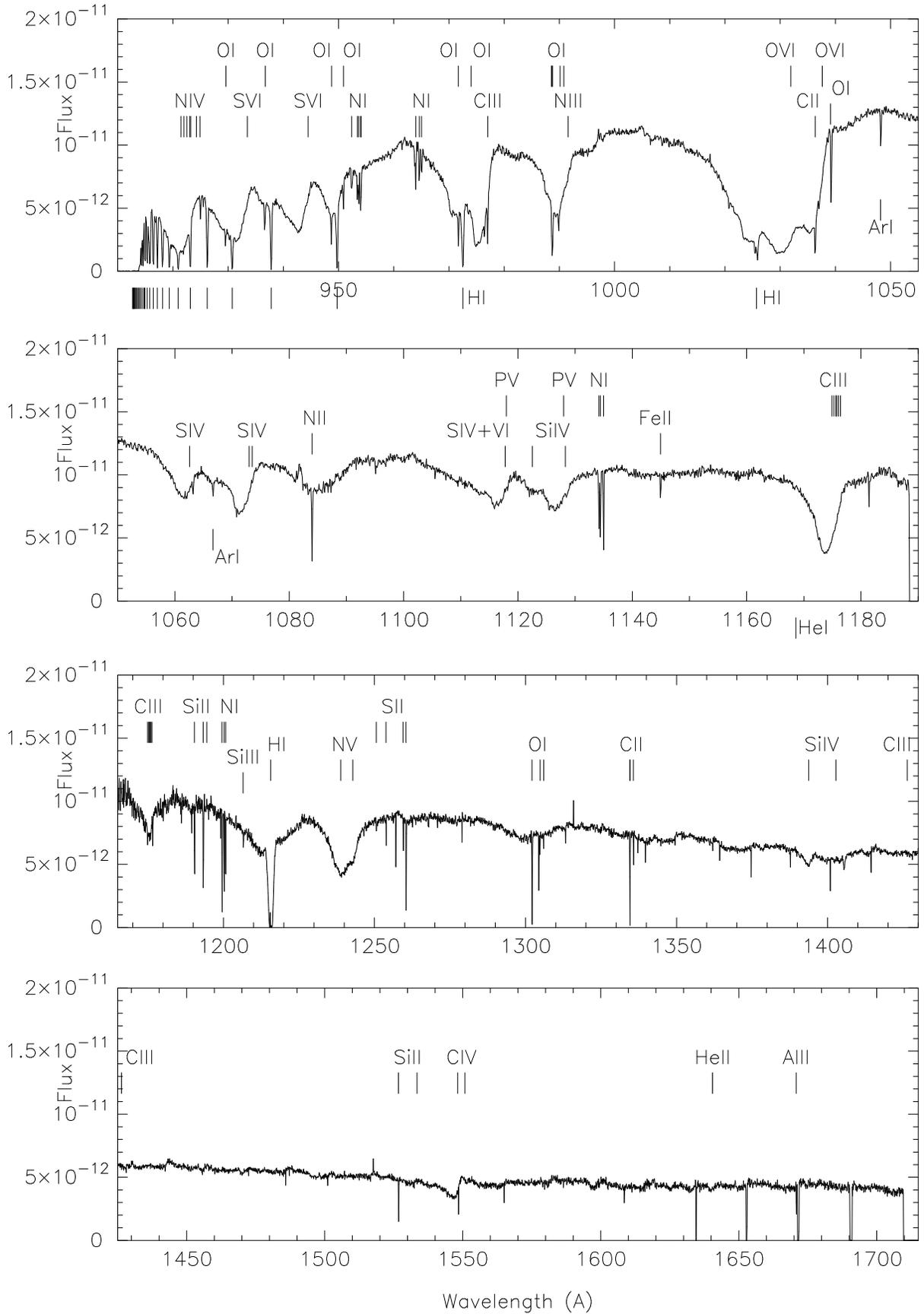}
\epsscale{0.80}
\figcaption{
$FUSE$ plus $STIS$ combined spectra with line identifications.
\label{f1}}
\end{figure}
%%%%%%%%%%%%%%%%%%%%%%%%%%%%%%%%%%%%%%%%%%%%%%%%%%%%%%%%%%%%%%%%%%%

%%%%%%%%%%%%%%%%%%%%%%%%%%%%%%%%%%%%%%%%%%%%%%%%%%%%%%%%%%%%%%%%%%%
\begin{figure}[tb]
\epsscale{0.97}
\plotone{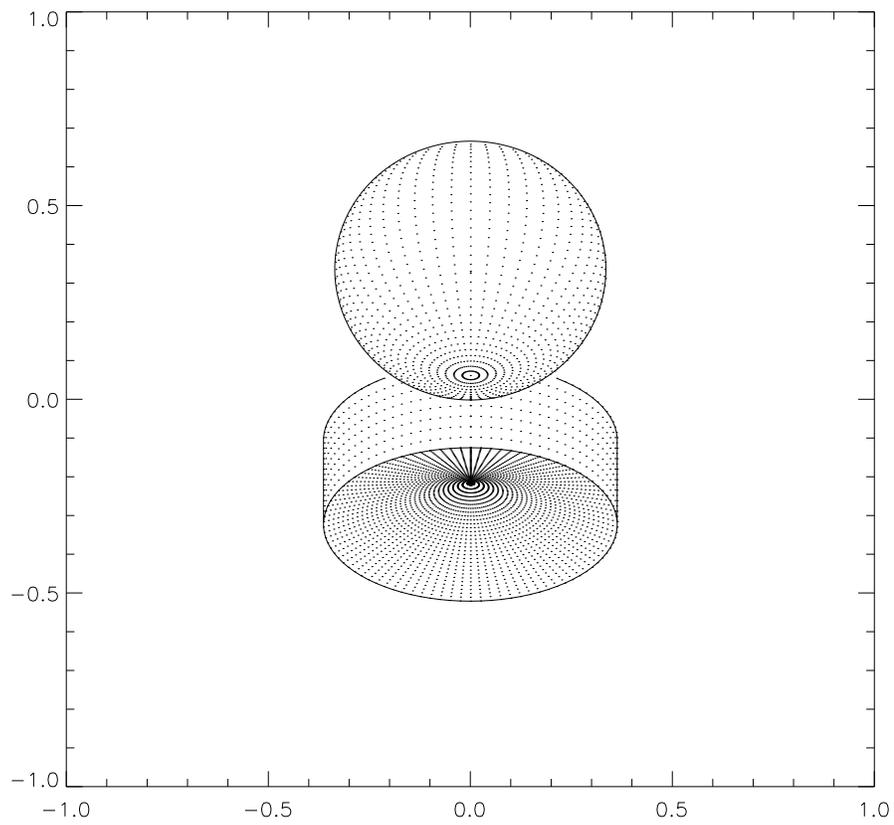}
\epsscale{1.00}
\figcaption{IX Vel model projected on the plane of the sky. The
orbital inclination is $57\arcdeg$ and the orbital phase is 0.0.
\label{f2}}
\end{figure}
%%%%%%%%%%%%%%%%%%%%%%%%%%%%%%%%%%%%%%%%%%%%%%%%%%%%%%%%%%%%%%%%%%%

%%%%%%%%%%%%%%%%%%%%%%%%%%%%%%%%%%%%%%%%%%%%%%%%%%%%%%%%%%%%%%%%%%%
\begin{figure}[tb]
\epsscale{0.97}
\plotone{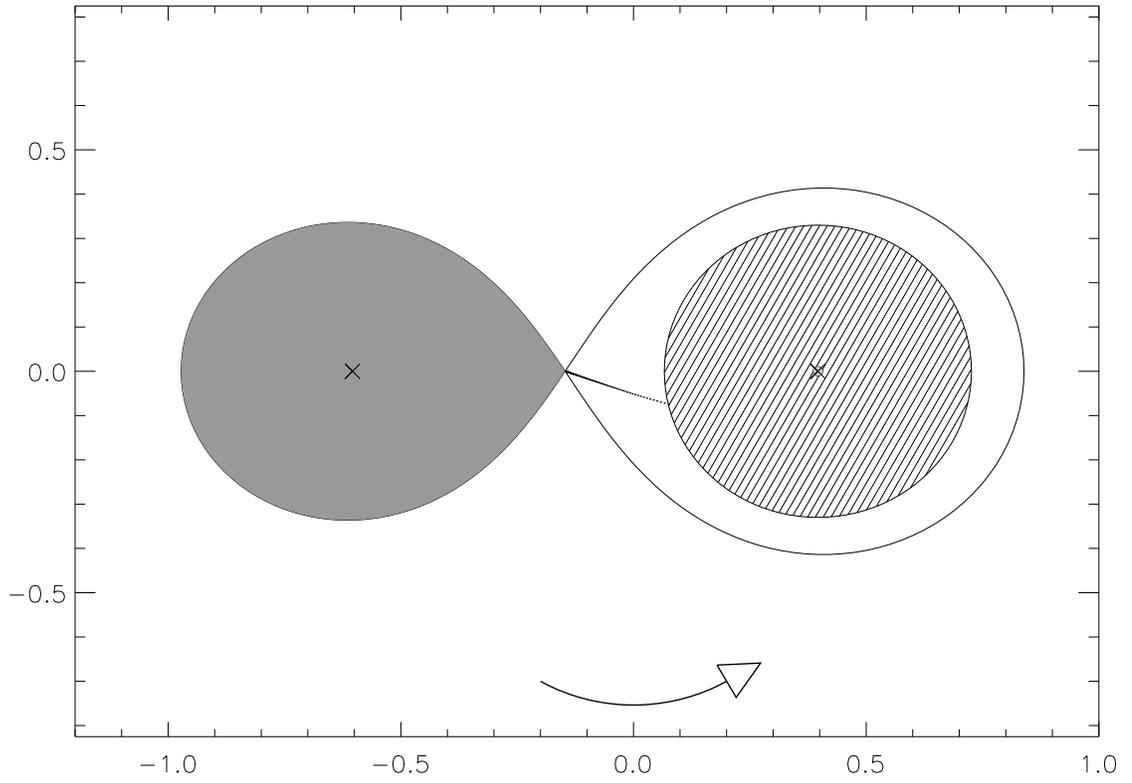}
\epsscale{1.00}
\figcaption{
Plan view of IX Vel system. The mass stream trajectory follows the theory
of \citet{ls1975}. See the text for a discussion.
\label{f3}}
\end{figure}
%%%%%%%%%%%%%%%%%%%%%%%%%%%%%%%%%%%%%%%%%%%%%%%%%%%%%%%%%%%%%%%%%%%

%%%%%%%%%%%%%%%%%%%%%%%%%%%%%%%%%%%%%%%%%%%%%%%%%%%%%%%%%%%%%%%%%%%
\begin{figure}[tb]
\epsscale{0.97}
\plotone{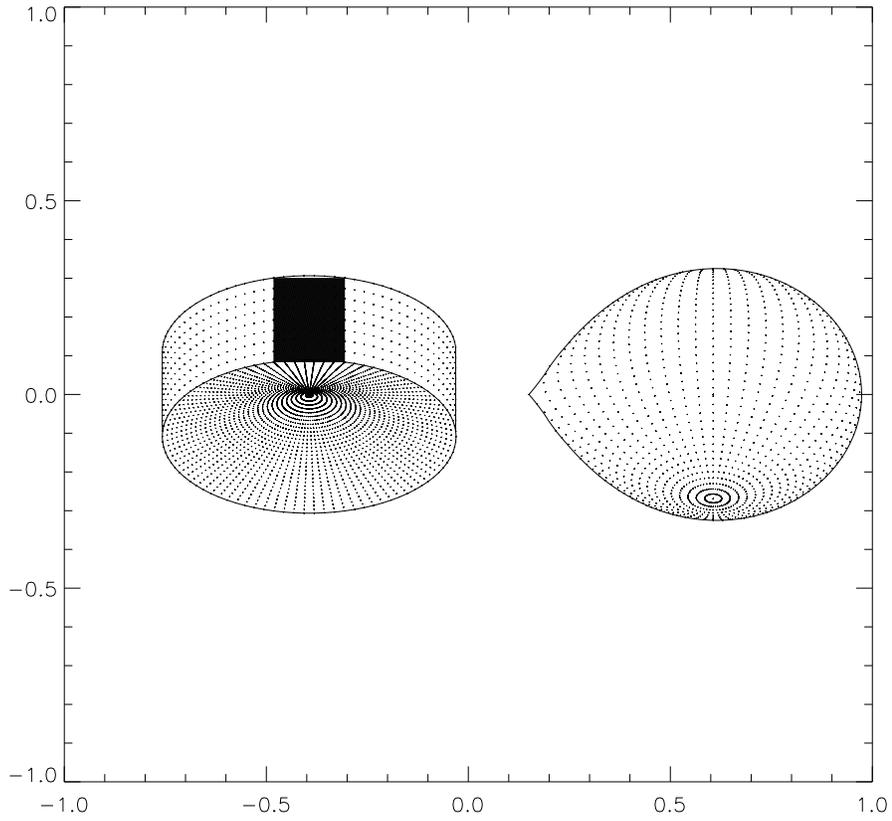}
\epsscale{1.00}
\figcaption{IX Vel model at orbital phase 0.75. The dark rim region
marks the bright spot, not shown on Figure~2. The bright
region is downwind from the point of impact of the mass transfer stream.
\label{f4}}
\end{figure}
%%%%%%%%%%%%%%%%%%%%%%%%%%%%%%%%%%%%%%%%%%%%%%%%%%%%%%%%%%%%%%%%%%%

%%%%%%%%%%%%%%%%%%%%%%%%%%%%%%%%%%%%%%%%%%%%%%%%%%%%%%%%%%%%%%%%%%%
\begin{figure}[tb]
\epsscale{0.97}
\plotone{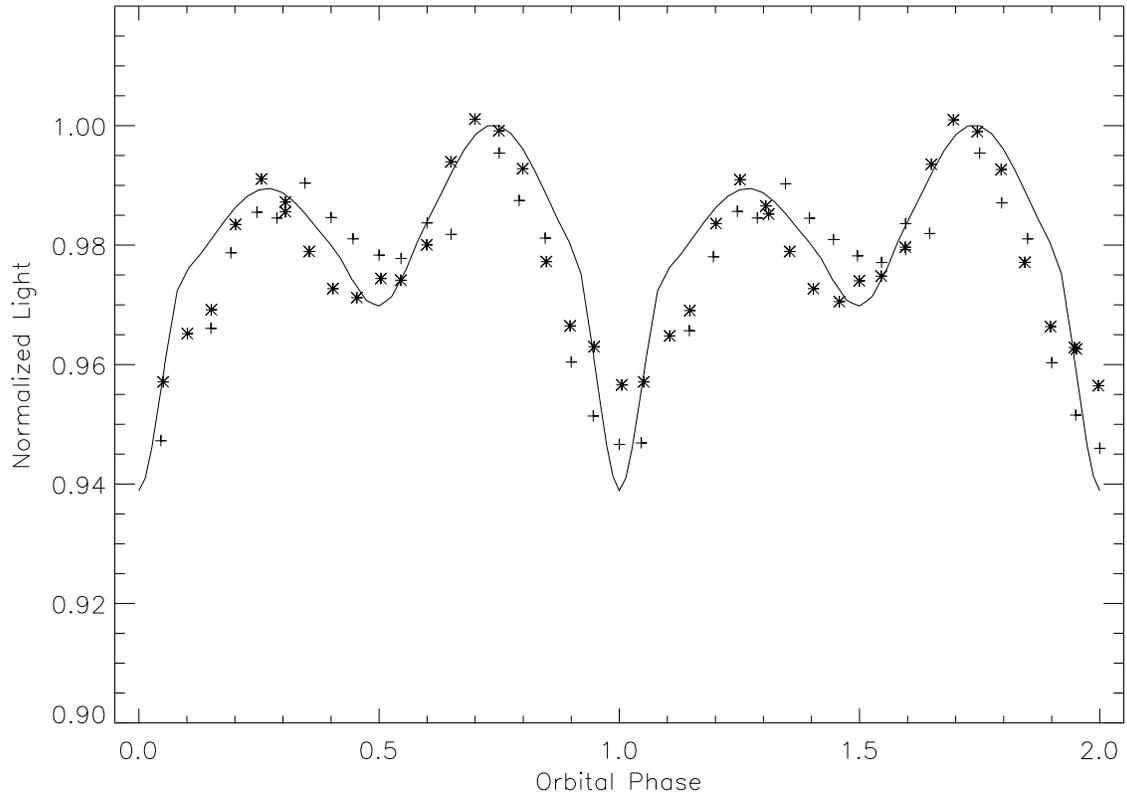}
\epsscale{1.00}
\figcaption{Synthetic light curve and observed (binned) K band observations
for an orbital inclination of $57\arcdeg$.
The crosses mark observations of 1984 January 16-23; the asterisks mark
observations of 1987 January 12-16.
See the text for details.
\label{f5}}
\end{figure}
%%%%%%%%%%%%%%%%%%%%%%%%%%%%%%%%%%%%%%%%%%%%%%%%%%%%%%%%%%%%%%%%%%%

\clearpage

%%%%%%%%%%%%%%%%%%%%%%%%%%%%%%%%%%%%%%%%%%%%%%%%%%%%%%%%%%%%%%%%%%%
\begin{figure}[tb]
\epsscale{0.97}
\plotone{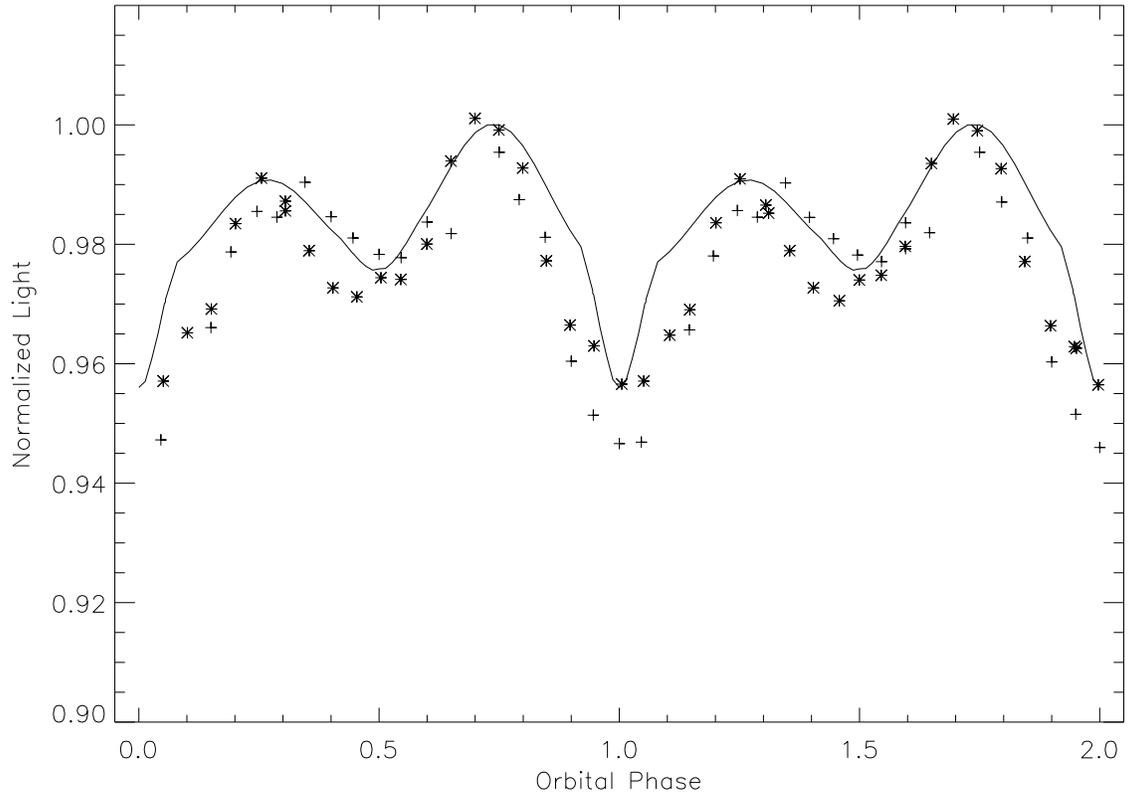}
\epsscale{1.00}
\figcaption{As in Figure~5 but for an orbital inclination of $54\arcdeg$.
The light curve amplitude is too small.
\label{f6}}
\end{figure}
%%%%%%%%%%%%%%%%%%%%%%%%%%%%%%%%%%%%%%%%%%%%%%%%%%%%%%%%%%%%%%%%%%%

%%%%%%%%%%%%%%%%%%%%%%%%%%%%%%%%%%%%%%%%%%%%%%%%%%%%%%%%%%%%%%%%%%%
\begin{figure}[tb]
\epsscale{0.97}
\plotone{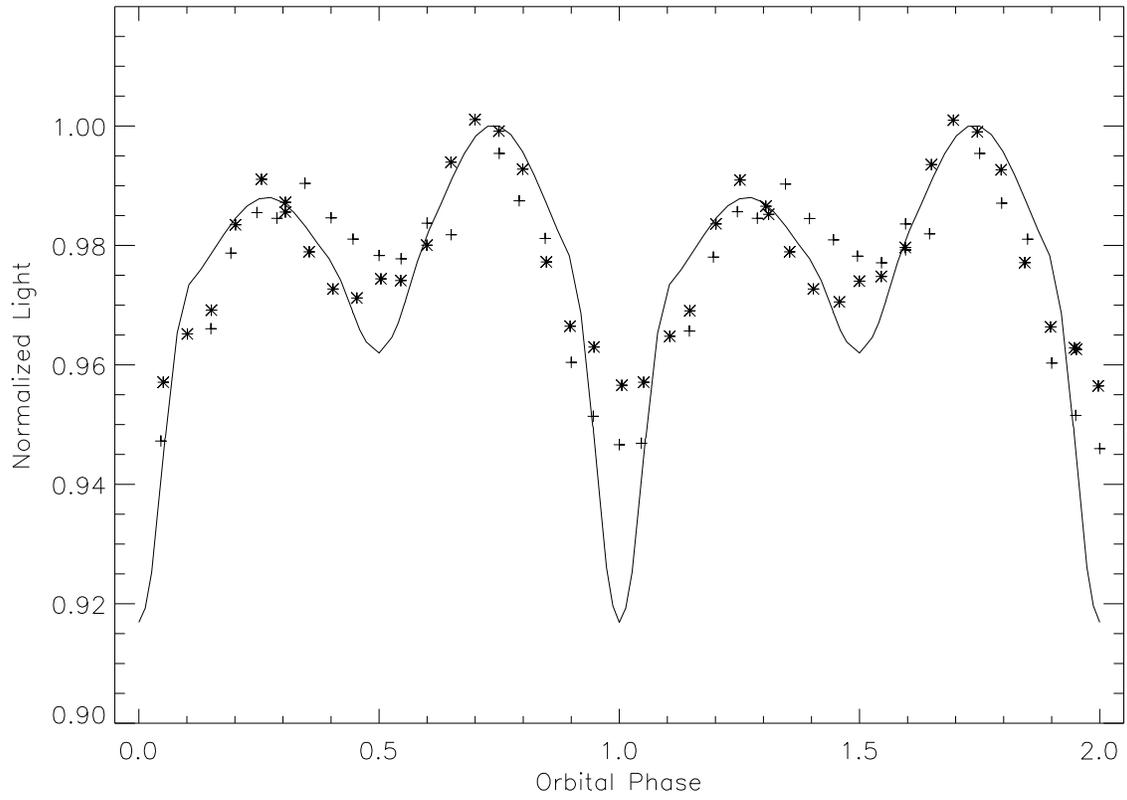}
\epsscale{1.00}
\figcaption{As in Figure~5 but for an orbital inclination of $60\arcdeg$.
The light curve amplitude is too large. See the text for a discussion.
\label{f7}}
\end{figure}
%%%%%%%%%%%%%%%%%%%%%%%%%%%%%%%%%%%%%%%%%%%%%%%%%%%%%%%%%%%%%%%%%%%

%%%%%%%%%%%%%%%%%%%%%%%%%%%%%%%%%%%%%%%%%%%%%%%%%%%%%%%%%%%%%%%%%%%
\begin{figure}[tb]
\epsscale{0.97}
\plotone{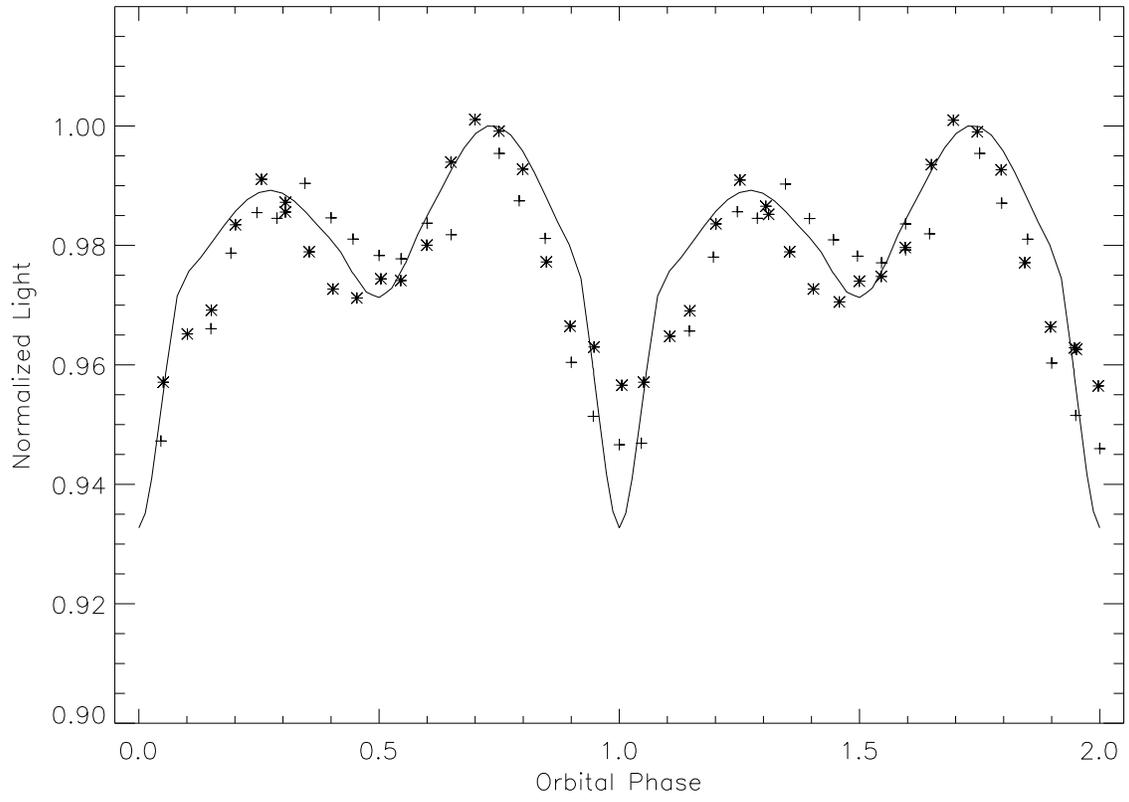}
\epsscale{1.00}
\figcaption{
As in Figure~5 but for the final system parameters. See the text for a discussion.
\label{f8}}
\end{figure}
%%%%%%%%%%%%%%%%%%%%%%%%%%%%%%%%%%%%%%%%%%%%%%%%%%%%%%%%%%%%%%%%%%%

%%%%%%%%%%%%%%%%%%%%%%%%%%%%%%%%%%%%%%%%%%%%%%%%%%%%%%%%%%%%%%%%%%%
\begin{figure}[tb]
\epsscale{0.97}
\plotone{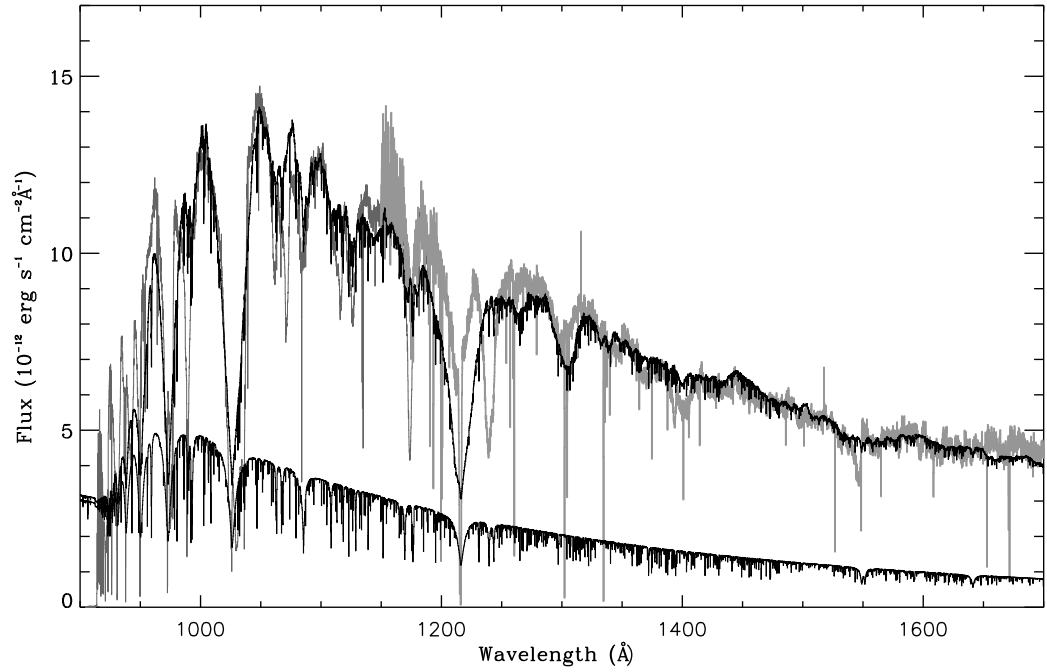}
\epsscale{1.00}
\figcaption{Synthetic system spectrum superposed on observed spectra.
The light gray plot marks the $FUSE$ spectrum; the darker gray plot marks the $STIS$
spectrum.
The contribution of a 60,000K WD is at the bottom.
\label{f9}}
\end{figure}
%%%%%%%%%%%%%%%%%%%%%%%%%%%%%%%%%%%%%%%%%%%%%%%%%%%%%%%%%%%%%%%%%%%

%%%%%%%%%%%%%%%%%%%%%%%%%%%%%%%%%%%%%%%%%%%%%%%%%%%%%%%%%%%%%%%%%%%
\begin{figure}[tb]
\epsscale{0.97}
\plotone{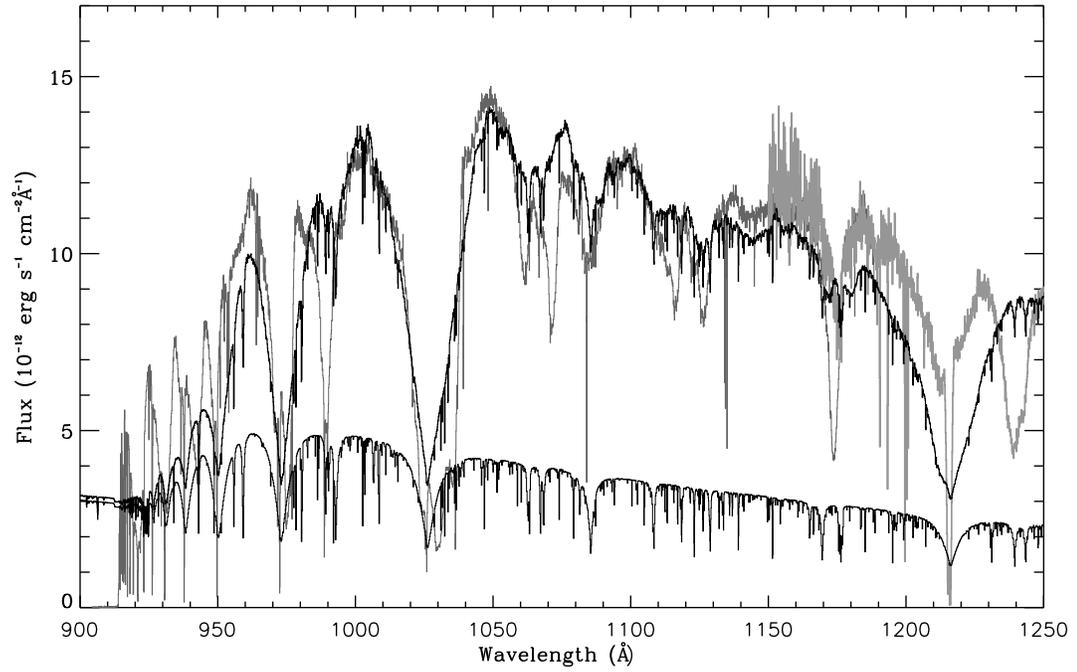}
\epsscale{1.00}
\figcaption{Detail of Figure~9.
Note the strong and unmodeled wind lines.
The contribution of a 60,000K WD is at the bottom.
\label{f10}}
\end{figure}
%%%%%%%%%%%%%%%%%%%%%%%%%%%%%%%%%%%%%%%%%%%%%%%%%%%%%%%%%%%%%%%%%%%

%%%%%%%%%%%%%%%%%%%%%%%%%%%%%%%%%%%%%%%%%%%%%%%%%%%%%%%%%%%%%%%%%%%
\begin{figure}[tb]
\epsscale{0.97}
\plotone{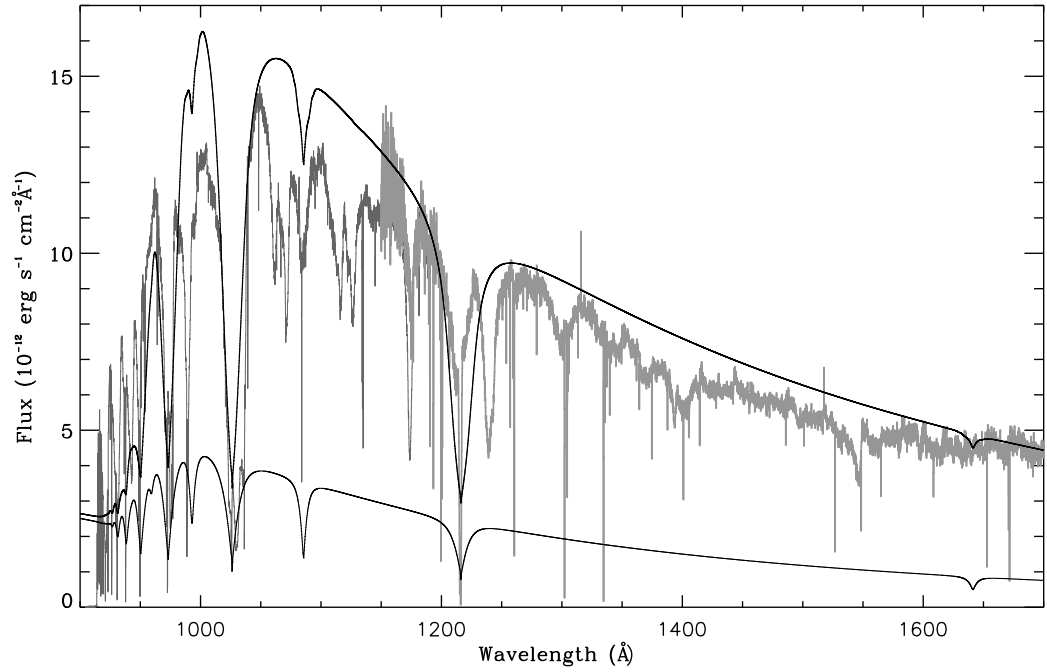}
\epsscale{1.00}
\figcaption{Synthetic system continuum spectrum plus H I and He II lines
superposed on the observed spectra.
Compare with Figure~9: Note that inclusion of line opacity (Figure~9) is necessary
for a satisfactory fit.
\label{f11}}
\end{figure}
%%%%%%%%%%%%%%%%%%%%%%%%%%%%%%%%%%%%%%%%%%%%%%%%%%%%%%%%%%%%%%%%%%%

%%%%%%%%%%%%%%%%%%%%%%%%%%%%%%%%%%%%%%%%%%%%%%%%%%%%%%%%%%%%%%%%%%%
\begin{figure}[tb]
\epsscale{0.97}
\plotone{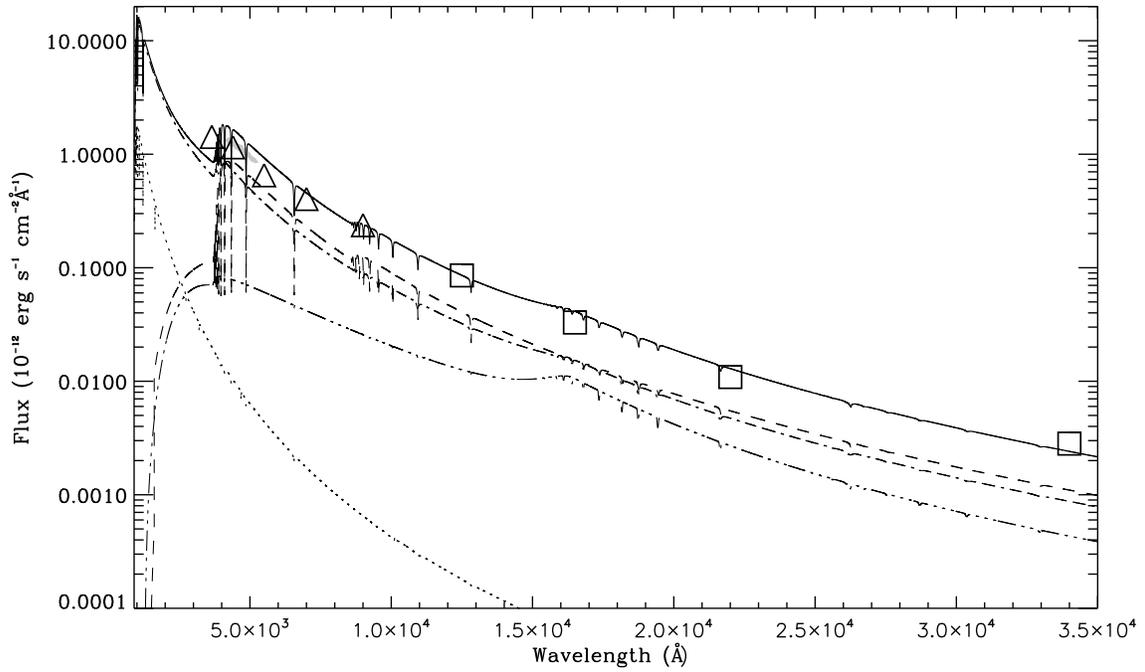}
\epsscale{1.00}
\figcaption{Plots of separate contributors to the system synthetic spectrum compared with
photometric data. All synthetic spectra are theoretical continuum plus H I and He II lines.
The continuous curve is the system synthetic spectrum; the dotted line is
the WD; the dashed line is the secondary star; the dash dot line is the accretion disk face;
the dash dot dot dot line is the accretion disk rim.
The triangles represent UBVRI photometry by \citet{gar1984} and squares represent JHKL
photometry by \citet{haug1988} and \citet{war1984}.
\label{f12}}
\end{figure}
%%%%%%%%%%%%%%%%%%%%%%%%%%%%%%%%%%%%%%%%%%%%%%%%%%%%%%%%%%%%%%%%%%%

%%%%%%%%%%%%%%%%%%%%%%%%%%%%%%%%%%%%%%%%%%%%%%%%%%%%%%%%%%%%%%%%%%%
\begin{figure}[tb]
\epsscale{0.97}
\plotone{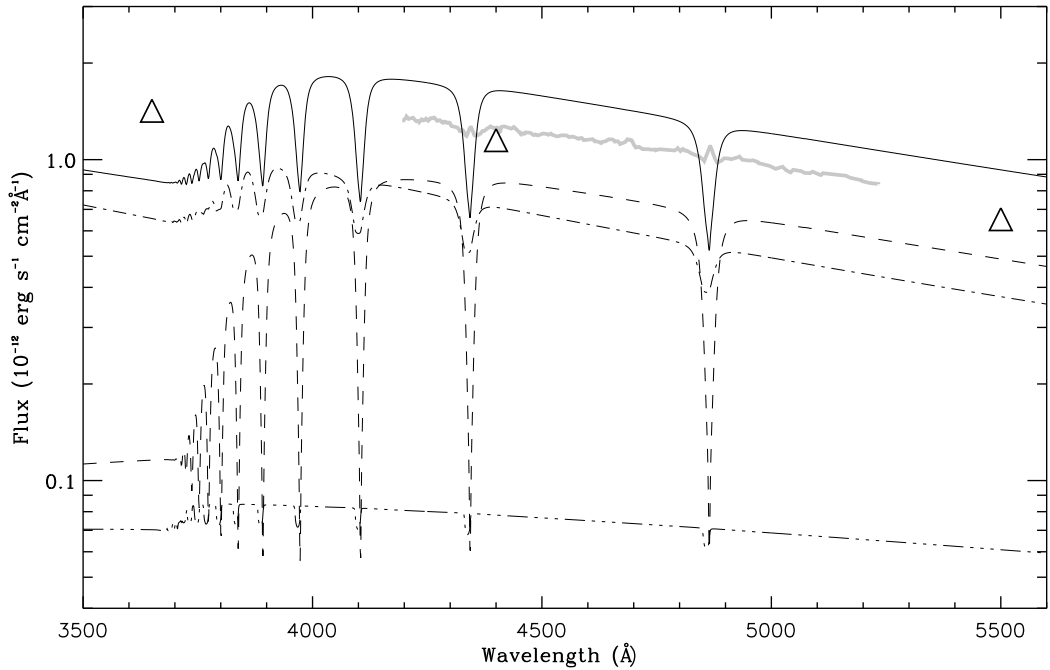}
\epsscale{1.00}
\figcaption{Detail of Figure~12. Based on the continuum spectra, the secondary star 
appears to be the major contributor to the
line strengths of the system Balmer lines. The orbital phase is 0.25, with the secondary star
receding. Note the slight redward shift of the secondary star lines relative to the face and
rim of the accretion disk lines. The lines of the latter two objects are rotationally broadened.
The gray line, also appearing in Figure~12, is an optical spectrum by \citet{war1983}.
\label{f13}}
\end{figure}
%%%%%%%%%%%%%%%%%%%%%%%%%%%%%%%%%%%%%%%%%%%%%%%%%%%%%%%%%%%%%%%%%%%

%%%%%%%%%%%%%%%%%%%%%%%%%%%%%%%%%%%%%%%%%%%%%%%%%%%%%%%%%%%%%%%%%%%
\begin{figure}[tb]
\epsscale{0.97}
\plotone{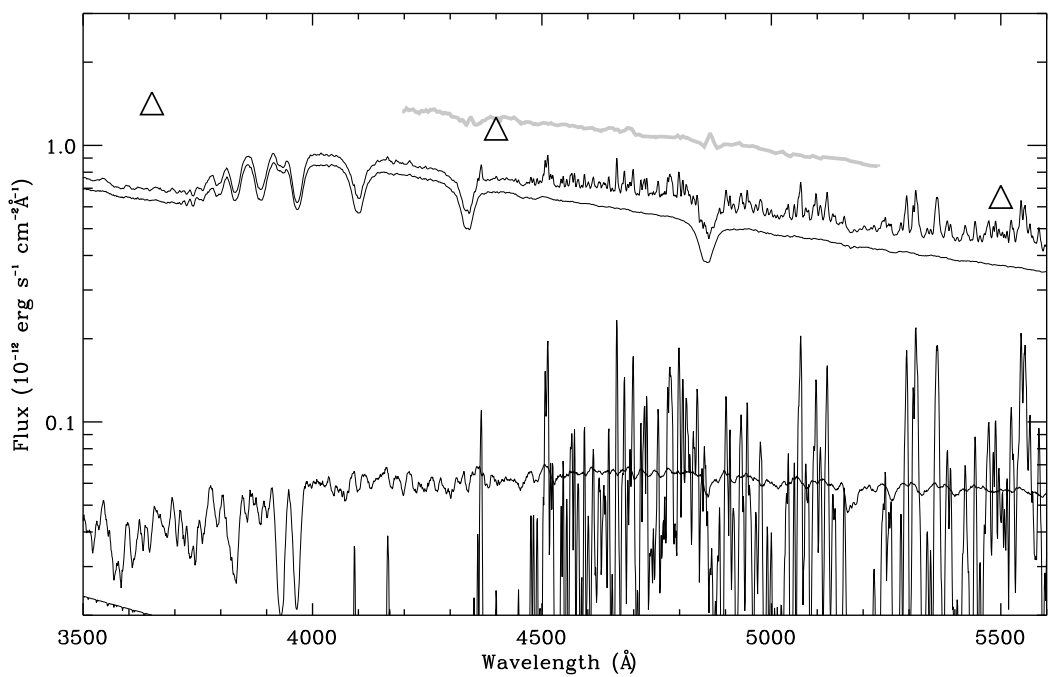}
\epsscale{1.00}
\figcaption{
Line synthetic spectrum corresponding to Figure~13.	Observational data as in Figure~12.
The accretion disk face synthetic spectrum lies immediately below the system spectrum.
The rim synthetic spectrum runs roughly horizontally across the plot, and the secondary
star synthetic spectrum shows in the lower left corner and continues as the very jagged plot.
This line spectrum shows that the accretion disk face actually is the major contributor to the
Balmer lines. Compare with Figure~13; note the major change in the contribution of the 
secondary star. See the text for a discussion.
\label{f14}}
\end{figure}
%%%%%%%%%%%%%%%%%%%%%%%%%%%%%%%%%%%%%%%%%%%%%%%%%%%%%%%%%%%%%%%%%%%

%%%%%%%%%%%%%%%%%%%%%%%%%%%%%%%%%%%%%%%%%%%%%%%%%%%%%%%%%%%%%%%%%%%
\begin{figure}[tb]
\epsscale{0.97}
\plotone{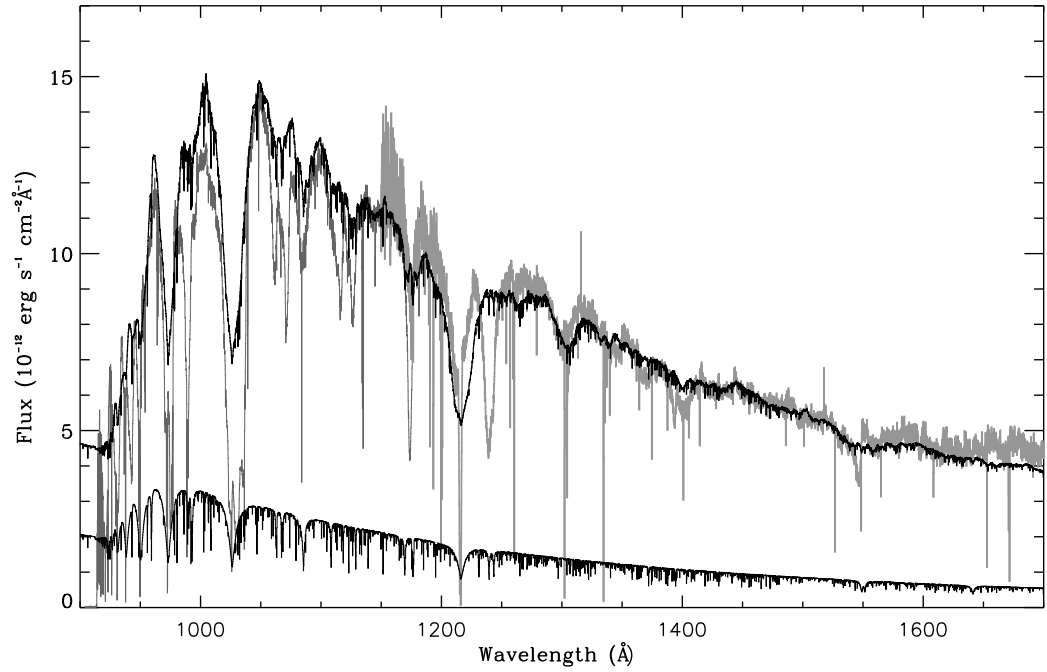}
\epsscale{1.00}
\figcaption{
Standard model accretion disk for $\dot{M}=5.0{\times}10^{-9}~{M}_{\odot}{\rm yr}^{-1}$ compared
with $FUSE$ and $STIS$ spectra.  
See the text
for a discussion.
\label{f15}}
\end{figure}
%%%%%%%%%%%%%%%%%%%%%%%%%%%%%%%%%%%%%%%%%%%%%%%%%%%%%%%%%%%%%%%%%%%

\end{document}